\documentclass[12pt]{article}
\pdfoutput=1

\usepackage[utf8]{inputenc}
\usepackage[left=2.55cm, right=2.55cm, top=2.55cm, bottom=2.55cm]{geometry}
\usepackage{amsmath,amssymb,amsbsy}
\usepackage{slashed}
\usepackage{xcolor}
\usepackage{graphicx}
\usepackage{url}
\usepackage{cancel}
\usepackage{cite}
\usepackage[colorlinks=true,allcolors=darkpurple,pdfborder={0 0 0},linktocpage=false]{hyperref}
\usepackage{tabularx,booktabs}
\usepackage{multicol}
\usepackage{units}
\usepackage{xspace}
\usepackage[labelfont=bf]{caption}
\usepackage[section]{placeins}
\usepackage{enumitem}
\usepackage{ulem}
\usepackage{subcaption}

\definecolor{darkred}{rgb}{0.6,0,0}
\definecolor{darkpurple}{rgb}{0.5,0,0.5}
\def\vev#1{\left\langle #1\right\rangle}
\def\L{\mathcal{L}}
\def\M{\mathcal{M}}
\def\O{\mathcal{O}}
\def\hc{\text{h.c.}}
\def\BR{\text{BR}}

\def\z2{$\mathbb{Z}_2$}
\def\Tr{\text{Tr}}

\def\id{\mathbb{I}}
\def\one{$\mathbf{1}$}
\def\two{$\mathbf{2}$}

\def\U1L{$\mathrm{U(1)}_L$}

\definecolor{avblue}{rgb}{0.0, 0.0, 0.8}
\definecolor{asparagus}{rgb}{0.53, 0.66, 0.42}
\definecolor{aqua}{rgb}{0.4, 0.6, 0.7}

\newcommand{\AddrIFIC}{%
  Instituto de F\'{i}sica Corpuscular, CSIC-Universitat de Val\`{e}ncia, 46980 Paterna, Spain}

\newcommand{\AddrFISTEO}{%
  Departament de F\'{\i}sica Te\`{o}rica, Universitat de Val\`{e}ncia, 46100 Burjassot, Spain}

\begin{document}

\vspace*{-2cm}
\begin{flushright}
IFIC/23-49 \\
\vspace*{2mm}
%\today
\end{flushright}

\begin{center}
\vspace*{15mm}

\vspace{1cm}
{\Large \bf 
The majoron coupling to charged leptons
} \\
\vspace{1cm}

{\bf Antonio Herrero-Brocal$^{\text{a}}$, Avelino Vicente$^{\text{a,b}}$}

 \vspace*{.5cm} 
 $^{(\text{a})}$ \AddrIFIC \\\vspace*{.2cm} 
 $^{(\text{b})}$ \AddrFISTEO

 \vspace*{.3cm}
\href{mailto:antonio.herrero@ific.uv.es}{antonio.herrero@ific.uv.es},
\href{mailto:avelino.vicente@ific.uv.es}{avelino.vicente@ific.uv.es}
\end{center}

\vspace*{10mm}
\begin{abstract}\noindent\normalsize
The particle spectrum of all Majorana neutrino mass models with
spontaneous violation of global lepton number include a Goldstone
boson, the so-called majoron. The presence of this massless
pseudoscalar changes the phenomenology dramatically. In this work we
derive general analytical expressions for the 1-loop coupling of the
majoron to charged leptons. These can be applied to any model
featuring a majoron that have a clear hierarchy of energy scales,
required for an expansion in powers of the low-energy scale to be
valid. We show how to use our general results by applying them to some
example models, finding full agreement with previous results in
several popular scenarios and deriving novel ones in other setups.
\end{abstract}

\section{Introduction}
\label{sec:intro}

Lepton number is an accidental symmetry in the Standard Model (SM) of
particle physics. Many extensions of this minimal model actually
include sources of lepton number violation. This is the case of all
Majorana neutrino mass models, in which the \U1L lepton number
symmetry is broken. When this symmetry is global and its breaking is
spontaneous, a massless Goldstone boson arises in the spectrum, the
majoron
($J$)~\cite{Chikashige:1980qk,Chikashige:1980ui,Schechter:1981cv,Gelmini:1980re,Aulakh:1982yn}. This
massless pseudoscalar has dramatic phenomenological implications and
can be probed by many experiments, including the search for rare
low-energy processes~\cite{Escribano:2020wua} and invisible Higgs
decays at high-energy colliders~\cite{Joshipura:1992hp}, as well as
due to its impact on many cosmological
observables~\cite{AristizabalSierra:2014uzi,Bonilla:2019ipe,Escudero:2019gvw,DeRomeri:2022cem}.

The flavor structure of the majoron couplings to charged leptons is
crucial for the phenomenology~\cite{Cheng:2020rla,Sun:2021jpw}. As
explained below, flavor diagonal couplings are strongly constrained by
astrophysical observations, while the flavor off-diagonal ones may
induce exotic processes like $\ell_\alpha \to \ell_\beta \, J$, with
$\alpha \neq \beta$. Different models also induce majoron couplings to
charged leptons at different loop orders. Tree-level diagonal majoron
couplings to charged leptons are actually induced in many scenarios
beyond the SM. For instance, in models that generate them via mixing
between the Higgs and the singlet that breaks lepton number
spontaneously. However, in this case the couplings turn out to be
diagonal in the charged lepton mass basis. An alternative mechanism
must be introduced in order to generate tree-level off-diagonal
couplings. For instance, these are induced if the majoron couples
directly to the charged leptons or to other fermions that mix with
them~\cite{Escribano:2021uhf}.

We study the 1-loop coupling of the majoron to a pair of charged
leptons. In contrast to previous works that focus on specific models,
typically the type-I seesaw with spontaneous lepton number
violation~\cite{Chikashige:1980ui,Pilaftsis:1993af,Garcia-Cely:2017oco,Heeck:2019guh},
we derive general expressions valid for (virtually) any model. We
consider all 1-loop diagrams leading to a majoron coupling to charged
leptons and expand the resulting analytical expressions in powers of
the light neutrino masses (or other related low-energy scale). The
main result of our work is a set of formulae that can be readily
applied to any majoron model of interest to obtain the couplings to
charged leptons. In order to demonstrate their use, we illustrate the
application of our analytical results to several example models. This
allows us to recover well-known results in some popular scenarios,
which constitutes a non-trivial cross-check of our calculation. We
also obtain novel results in other less studied models.

The rest of the manuscript is organized as follows. We present our
setup, discuss the current bounds on the majoron couplings to charged
leptons and derive general expressions for them in
Sec.~\ref{sec:coup}. These general results are applied to specific
models in Sec.~\ref{sec:examples}. This allows us to show how to use
our analytical results. We summarize our work and discuss further
directions in Sec.~\ref{sec:sum}. Finally, appendices~\ref{sec:app1}
and \ref{sec:app2} contain additional details about our results.

\section{The majoron coupling to charged leptons}
\label{sec:coup}

The Lagrangian describing the interaction of a majoron with a pair of
charged leptons can be generally written as~\cite{Escribano:2020wua},
\begin{equation} \label{eq:llJ}
\mathcal{L}_{\ell \ell J} = J \, \bar{\ell}_\beta \left( S_L^{\beta \alpha} \, P_L + S_R^{\beta \alpha} \, P_R \right) \ell_{\alpha} + \hc = J \, \bar{\ell}_\beta \left[ S^{\beta \alpha} \, P_L + \left( S^{\alpha \beta} \right)^* \, P_R \right] \ell_{\alpha} \, .
\end{equation}
Here $P_{L,R} = \frac{1}{2} \left( 1 \mp \gamma_5 \right)$ are the
usual chiral projectors while $\ell_{\alpha,\beta}$ are the charged
leptons, with $\alpha, \beta = 1,2,3$ two flavor indices, and
\begin{equation}
S^{\beta \alpha} = S_L^{\beta \alpha} +\left( S_R^{ \alpha \beta} \right)^*  \, .
\end{equation}
All flavor combinations are considered: $\beta\alpha = \left\{
ee,\mu\mu,\tau\tau,e\mu,e\tau,\mu\tau\right\}$. We note that majorons
are pseudoscalar states. This implies that the diagonal $S^{\beta
  \beta} = S_L^{\beta \beta} + \left(S_R^{\beta \beta} \right)^*$
couplings are purely imaginary at all orders in perturbation theory~\cite{Ecker:1983qf}. We are interested
in models that induce flavor off-diagonal couplings at the 1-loop
level. Therefore, while flavor diagonal couplings will be allowed at
tree-level, the off-diagonal ones will be absent and only generated at
1-loop. In models that generate off-diagonal couplings at tree-level,
the 1-loop contributions considered in our work can be regarded as
just corrections, hence expected to be subdominant. For similar
  reasons, we will consider only diagonal tree-level couplings with
  quarks.

There are stringent constraints on both diagonal and off-diagonal
majoron couplings to charged leptons~\cite{Escribano:2020wua}. The
former are constrained by astrophysical observations. If produced,
majorons would escape astrophysical scenarios without interacting with
the medium, thus leading to a very efficient energy loss
mechanism. This has been studied in several recent
works~\cite{DiLuzio:2020wdo,Bollig:2020xdr,Calibbi:2020jvd,Croon:2020lrf,DiLuzio:2020jjp,Caputo:2021rux},
which have derived the limits
\begin{equation} \label{eq:See}
  \left| \text{Im} \, S^{e e} \right| < 2.1 \times 10^{-13}
\end{equation}
and
\begin{equation} \label{eq:Smumu}
 \left| \text{Im} \, S^{\mu \mu} \right| < 3.1 \times 10^{-9} \, .
\end{equation}
The off-diagonal majoron couplings to charged leptons also receive
strong constraints, in this case from searches of the flavor violating
decays $\ell_\alpha \to \ell_\beta \, J$. In order to specify the
bound, it proves convenient to define the combination
\begin{equation}
  \left| \widetilde S^{\beta \alpha} \right| = \left( \left| S^{\beta \alpha}_L \right|^2 + \left| S^{\beta \alpha}_R \right|^2 \right)^{1/2} \, ,
\end{equation}
that enters the $\ell_\alpha \to \ell_\beta \, J$ decay width as
\begin{equation} \label{eq:gamma}
\Gamma(\ell_\alpha \to \ell_\beta \, J) = \frac{m_\alpha}{32 \, \pi} \, \left| \widetilde S^{\beta \alpha} \right|^2 \, .
\end{equation}
The current limit on the branching ratio of $\mu^+ \to e^+ \, J$ was
obtained at TRIUMF~\cite{Jodidio:1986mz}. Taking into account all
possible chiral structures for the majoron coupling, one can estimate
the limit $\BR \left( \mu \to e \, J \right) \lesssim
10^{-5}$~\cite{Hirsch:2009ee}, which in turn implies
\begin{equation} \label{eq:mulim}
  \left| \widetilde S^{e \mu} \right| < 5.3 \times 10^{-11} \, .
\end{equation}
Finally, the currently best experimental limits on $\tau$ decays
including majorons were set by the Belle II
collaboration~\cite{Belle-II:2022heu}. They can be used to derive the
bounds
\begin{equation} \label{eq:taulim}
  \begin{split}
    & \left| \widetilde S^{e \tau} \right| < 3.5 \times 10^{-7} \, , \\
    & \left| \widetilde S^{\mu \tau} \right| < 2.7 \times 10^{-7} \, .
  \end{split}
\end{equation}

\subsection{General setup}
\label{subsec:setup}

Our goal is to obtain generic expressions for the majoron coupling to
a pair of charged leptons. We adopt a general setup with $3$ light
charged leptons, $\ell_\alpha = \left\{e,\mu,\tau\right\}$, $N$
neutral leptons, $n_i = \left\{ n_1, \dots , n_N \right\}$, out of
which at least $3$ must be light, and $6$ quarks, $q_i =
\left\{u,c,t,d,s,b\right\}$.~\footnote{In the following, we denote
the charged lepton flavor index with a Greek character, $\alpha =
1,2,3$, while the neutral fermion and quark flavor indices will
be denoted with a Latin character, $i = 1,\dots,N$ and $i =
1,\dots,6$, respectively.} Note that only the usual SM charged
  leptons and quarks are considered and that all quarks are
  generically denoted by $q_i$, without any distinction between up-
  and down-type quarks. We also assume the neutral leptons to be of
  Majorana nature, which implies that their mass matrix in the gauge
  basis is symmetric. We also introduce a massive scalar $\rho$, a
massive pseudoscalar $\sigma$, a singly charged scalar $\eta^\pm$, a
scalar leptoquark $X$ and a vector leptoquark $Y_\mu$. All these
fields are mass eigenstates. In case more than one massive scalar,
pseudoscalar, charged scalar or leptoquark (of any type) is
present in the model, our results can be easily adapted by summing
over all mass eigenstates. In addition, we note that at least one
pseudoscalar will always be part of the spectrum, namely the massless
majoron. We will consider the following general interaction
Lagrangian:
\begin{equation} \label{eq:lag}
\L = \L_J + \L_\rho + \L_\sigma + \L_\eta + \L_{\rm LQ} + \L_Z + \L_W + \L_S \, .
\end{equation}
The first term describes the interaction of the majoron with a pair of
neutrinos, a pair of charged leptons or a pair of quarks, and is given by~\footnote{It may seem \textit{exotic} to consider a majoron coupling to quarks, but this interaction is indeed induced at tree-level in models in which the scalar singlet responsible for lepton number violation gets a non-zero mixing with the Higgs doublet after the electroweak and lepton number symmetries are spontaneously broken.}
\begin{equation} \label{eq:lagJ}
  \L_J =
  \bar{n}_i \left( A_{ij} P_R + A_{ji}^* P_L \right) \, n_j \, J
  + \bar{\ell}_\alpha \, K_{\alpha \alpha} \gamma_5 \, \ell_\alpha \, J + \bar{q}_i \, I_{i i} \gamma_5 \, q_i \, J \, .
\end{equation}
As already explained, we have considered only flavor-diagonal
couplings between the majoron and charged leptons and between the majoron and quarks (here, $K$ and $I$ are
purely imaginary diagonal matrices). The Yukawa interactions of the
scalar $\rho$ and the pseudoscalar $\sigma$ can be written as
\begin{align}
  \L_\rho &=
  \bar{\ell}_{\alpha} \left( G_{\alpha \beta} P_L + G_{\beta \alpha}^* P_R \right) \, \ell_\beta \, \rho \, , \label{eq:lagRho} \\
  \L_\sigma &=
  \bar{n}_i \left( B_{ij} P_R + B_{ji}^* P_L \right) \, n_j \, \sigma
  + \bar{\ell}_\alpha \left( C_{\alpha \beta} P_L + C_{\beta \alpha}^* P_R \right) \, \ell_\beta \, \sigma \, . \label{eq:lagSigma}
\end{align}
In principle, the scalar $\rho$ can also couple to a pair of
neutrinos, but we have decided to ignore these couplings, since they
would not contribute (at 1-loop) to the majoron coupling to charged
leptons. We also note that Eq.~\eqref{eq:lagSigma} reduces to
$\bar{\ell}_\alpha \, C_{\alpha \alpha}^* \gamma_5 \, \ell_\alpha \,
\sigma$, with $C_{\alpha \alpha}$ purely imaginary, when one considers
the interaction term between the pseudoscalar $\sigma$ and a pair of
charged leptons with the same flavor. Similarly, Eq.~\eqref{eq:lagRho}
reduces to $\bar{\ell}_\alpha \, G_{\alpha \alpha} \, \ell_\alpha \,
\sigma$, with $G_{\alpha \alpha}$ purely real, in the flavor diagonal
case. The charged scalar $\eta$ can also have a Yukawa coupling to a
charged lepton and a neutrino, given by
\begin{equation} \label{eq:lagEta}
  \L_\eta =
  \bar{\ell}_{\alpha} \left( D_L^{\alpha i} P_L + D_R^{\alpha i} P_R \right) \, n_i \, \eta + \hc \, .
\end{equation}
The $X$ and $Y$ leptoquarks couple to a quark and a charged lepton in the general form
\begin{equation}
 \L_{\rm LQ} =
  \bar{\ell}_{\alpha} \left( T_L^{\alpha i} P_L + T_R^{\alpha i} P_R \right) \, q_i \, X +  \bar{\ell}_{\alpha} Y_\mu \, \gamma^\mu \left( H_L^{\alpha i} \, P_L + H_R^{\alpha i} \, P_R \right) \, q_i \, + \hc \, .
\end{equation}
Since we denote up- and down-quarks as $q_i$ indiscriminately, the electric charges of $X$ and $Y$ are not fixed, but left as free parameters. However, these electric charges do not affect any of the results that follow and hence there is no need to distinguish between quark types.
$\L_Z$ and $\L_W$ contain the usual gauge bosons interactions with
neutrinos and charged leptons, which can be written
as~\cite{Pilaftsis:1991ug,Pilaftsis:1992st,Pilaftsis:1993af,Garcia-Cely:2017oco,Heeck:2019guh}
\begin{align}
    \mathcal{L}_Z &= \frac{g}{4 \, \cos \theta_W} \bar{n}_i \, \slashed{Z} \left( E_{ij} \, P_L - E_{ji} \, P_R \right) n_j +\frac{g}{ \, \cos \theta_W} \bar{\ell}_\alpha \, \slashed{Z} \left( v_\ell -a_\ell \gamma_5 \right) \ell_\alpha \, , \label{eq:lagZ} \\
    \mathcal{L}_W &=\frac{g}{\sqrt{2}} \left( \bar{\ell}_\alpha F_{\alpha i} \slashed{W}^- P_L \, n_i + \hc \right) \, , \label{eq:lagW}
\end{align}
where $E$ is an Hermitian matrix and $v_\ell$ and $a_\ell$ are the
usual $Z$-boson couplings to charged leptons. Finally, $ \L_S$
describes a cubic scalar interaction of the majoron with one scalar
and one pseudoscalar,
\begin{equation}
\L_S = \omega \, J \, \sigma \, \rho \, ,
\end{equation}
where $\omega$ is a real parameter with dimensions of mass. Sums over
charged lepton and neutrino flavor indices are implicitly assumed in
Eqs.~\eqref{eq:lagJ}-\eqref{eq:lagW}. The generic couplings $A$, $B$,
$C$, $D_{L,R}$, $E$, $F$ and $G_{L,R}$ have model-dependent
expressions in terms of the parameters of the specific model under
consideration. Nevertheless, they can generally be written as
\begin{align}
  A_{ij} &= U_{ki} \, \bar A_{kr} \, U_{rj} \, , \label{eq:rotA} \\
  B_{ij} &= U_{ki} \, \bar B_{kr} \, U_{rj} \, , \\
  D_{L}^{\alpha i} &= \bar D_{L}^{\alpha k} \, U_{ki}^* \, , \label{eq:rotDL} \\
  D_{R}^{\alpha i} &= \bar D_{R}^{\alpha k} \, U_{ki} \, , \label{eq:rotDR} \\
  E_{ij} &= U_{ki} \, \bar E_{kr} \, U_{rj}^*= \sum_{k=1}^3 \, U_{ki}  \, U_{kj}^*\, , \label{eq:rotE} \\
  F_{\alpha i} &= \bar F_{\alpha k} \, U_{ki}^*=U_{\alpha i}^* \, , \label{eq:rotF}
\end{align}
where $U$ is the unitary matrix implicitly defined by
\begin{align}
  M_{ij} &= U_{ki} \, \bar M_{kr} \, U_{rj} \, ,
\end{align}
with $M \equiv \text{diag} \left( m_{n_1} , \dots , m_{n_N} \right)$
and $\bar M$ the neutral lepton mass matrices in the mass and gauge
bases, respectively. We have assumed that we work in the charged
lepton mass basis. Therefore, $\bar A$, $\bar B$, $\bar D_{L,R}$,
$\bar E$ and $\bar F$ are the couplings in the gauge basis, while $C =
\bar C$ and $G = \bar G$. Eqs.~\eqref{eq:rotE} and \eqref{eq:rotF} are
completely general, since they just correspond to the transformation
of the usual neutral and charged currents, and the index $i=1,2,3$ in
these equations tags a gauge non-singlet. In the case of quark couplings we prefer to keep them in the mass basis, just for simplicity, since we are not interested in the gauge basis parameters.

\subsection{1-loop coupling}
\label{subsec:1loop}

\begin{figure}[th!]
  \centering
  \begin{subfigure}{0.42\linewidth}
    \includegraphics[width=\linewidth]{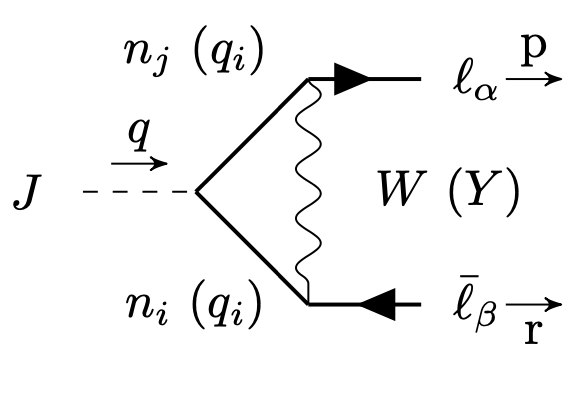}
    \caption{\textbf{$\boldsymbol{W}$ boson and $\boldsymbol{Y}$ contributions:} $\M_W + \M_{Y}$}
    \label{fig:Wdiag}
  \end{subfigure}
  \begin{subfigure}{0.57\linewidth}
    \includegraphics[width=\linewidth]{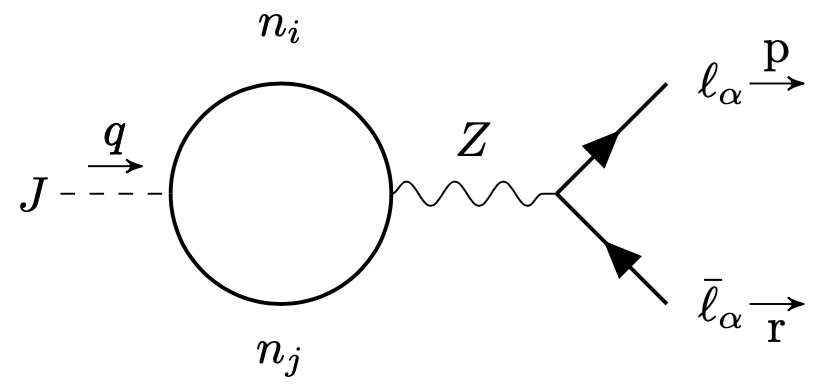}
    \caption{\textbf{$\boldsymbol{Z}$ boson contribution:} $\M_Z$}
    \label{fig:Zdiag}
  \end{subfigure}
  \vfill
  \bigskip
  \begin{subfigure}{0.42\linewidth}
    \includegraphics[width=\linewidth]{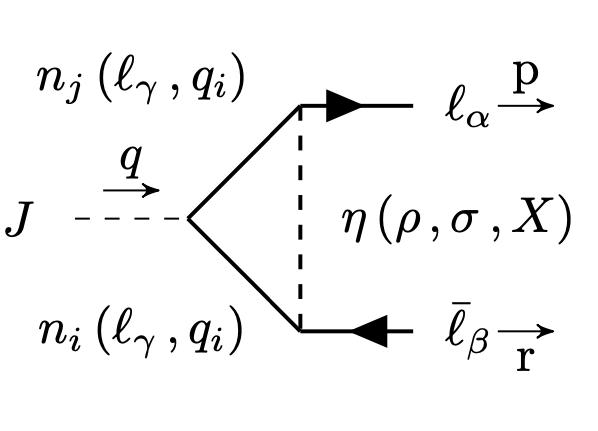}
    \caption{\textbf{$\boldsymbol{\eta}$, $\boldsymbol{\rho}$, $\boldsymbol{\sigma}$ and $\boldsymbol{X}$ contributions:} $\M_\eta + \M_\rho + \M_\sigma + \M_{X}$}
    \label{fig:EtaRhodiag}
  \end{subfigure}
  \begin{subfigure}{0.57\linewidth}
    \includegraphics[width=\linewidth]{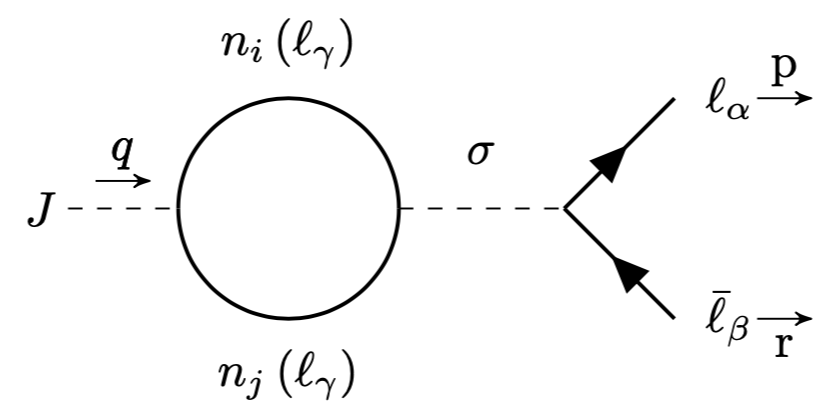}
    \caption{\textbf{$\boldsymbol{\sigma n}$ and $\boldsymbol{\sigma \ell}$ contributions:} $\M_{\sigma n} + \M_{\sigma\ell}$}
    \label{fig:Sigmadiag}
  \end{subfigure}
  \vfill
  \bigskip
  \begin{subfigure}{0.42\linewidth}
    \includegraphics[width=\linewidth]{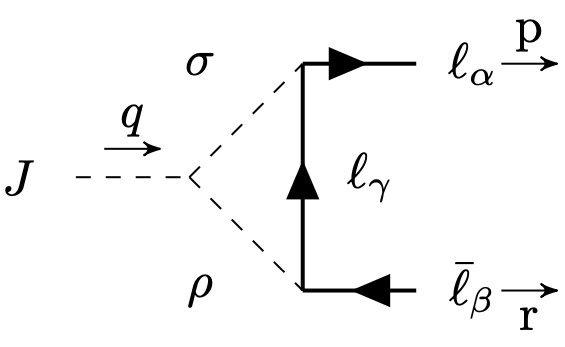}
    \caption{\textbf{$\boldsymbol{S}$ contribution:} $\M_S$}
    \label{fig:SigmaRhodiag}
  \end{subfigure}
  \caption{Feynman diagrams leading to the 1-loop coupling of the
    majoron to a pair of charged leptons.
    \label{fig:diagrams}}
\end{figure}

Let us now move on to the computation of the 1-loop coupling of the
majoron to a pair of charged leptons. First of all, we consider all
possible 1-loop diagrams leading to this coupling. They are shown in
Fig.~\ref{fig:diagrams}. Other 1-loop diagrams vanish due to the
pseudoscalar nature of the majoron. For instance, this would be the
case for the self-energy diagram in Fig.~\ref{fig:Sigmadiag} if we
replaced the pseudoscalar $\sigma$ by the scalar $\rho$. It is
important to notice a key aspect. In the scalar diagrams, the fermion
in the loop can either be a neutral or a charged lepton. We shall
consider both scenarios. We could in principle include charged leptons
in the $Z$ boson diagram. However, given that this diagram is always
flavor diagonal, such an inclusion would constitute a correction to
the tree-level majoron coupling, thus rendering it negligible. For
this reason we will not consider this option. If $\sigma$ couples
  to quarks, an additional $\sigma q$ contribution, analogous to
  $\sigma \ell$, would exist. The resulting $\M_{\sigma q}$ amplitude
  can be obtained by applying obvious $\ell \to q$ replacements to the
  $\M_{\sigma \ell}$ amplitude and we decided to omit it. Similarly,
  an additional $Z$ boson contribution with a quark closed loop can be
  added, again with a trivial $n \to q$ change. Finally, let us
  note that non-trivial electrically charged particles have been added
  to the spectrum. In such cases, we would have infinitely many
  different possibilities (due to the charge combinations). However,
  we emphasize once again that these electric charges do not affect
  our results.

The amplitudes of the Feynman diagrams in Fig.~\ref{fig:diagrams} will
be denoted by $\M_A$, with $A= \left\{ W, Z, \eta, \rho, \sigma, X, Y,
\sigma n, \sigma\ell, S \right\}$, and each amplitude corresponds to a
different diagram. For simplicity, we will work in the unitarity gauge, where the diagrams including SM Goldstone bosons are absent. Thus, the amplitudes are given by
\begin{align}
\M_W =& \, 2 \, \bar{v}_{\ell_\beta} \sum_{i,j} \int \frac{d^4k}{(2 \pi)^4} \frac{g}{\sqrt{2}} \, F_{\beta j} \, \gamma_\mu P_L  \frac{1}{k^2-M_W^2} \left(g^{\mu \nu} - \frac{k^\mu k^\nu}{M_W^2} \right)  \frac{\xout{k}-\xout{r}+m_j}{(k-r)^2-m_j^2} \nonumber \\
&
\left( A_{ij} P_R + A_{ji}^* P_L \right) \, \frac{\xout{k} +\xout{p} + m_i}{(k+p)^2-m_i^2} \frac{g}{\sqrt{2}} \, \gamma_\nu P_L \, F_{\alpha i}^* \, u_{\ell_\alpha} \, , \\
\M_Z =& -2\frac{g}{\cos \theta_W} \bar{v}_{\ell_\beta} \gamma_\mu \left(v_\ell - a_\ell \gamma_5 \right) \, u_{\ell_\alpha} \frac{1}{-M_Z^2} \left(g^{\mu \nu} - \frac{q^\mu q^\nu}{M_Z^2} \right) \nonumber \\
& \sum_{i,j}\int \frac{d^4k}{(2 \pi)^4} \frac{g}{4 \cos \theta_W} \, \Tr \left[ \gamma_\nu \left(E_{ij} \, P_L - E_{ji} \, P_R \right) \, \frac{\xout{k}+\xout{q}+m_j}{(k+q)^2-m_j^2} \left( A_{ij} P_R + A_{ji}^* P_L \right) \frac{\xout{k}+m_i}{k^2-m_i^2} \right] \, , \\
\M_\eta =& \, 2 \sum_{i,j} \bar{v}_{\ell_\beta} \int \frac{d^4k}{(2 \pi)^4} \, \left( D_L^{\beta j} P_L + D_R^{\beta j} P_R \right) \, \frac{\xout{k}-\xout{r}+m_j}{(k-r)^2-m_j^2} \left( A_{ij} P_R + A_{ji}^* P_L \right) \nonumber \\
& \frac{\xout{k} +\xout{p} + m_i}{(k+p)^2-m_i^2} \, \left[ \left(D_L^{\alpha i}\right)^* P_R + \left(D_R^{\alpha i}\right)^* P_L \right] \, \frac{1}{k^2-m_\eta^2} \, u_{\ell_\alpha} \, , \\
\M_\rho =& \,  \sum_{\gamma} \bar{v}_{\ell_\beta} \int \frac{d^4k}{(2 \pi)^4} \, \left( G_{\beta \gamma} P_L + G_{\gamma \beta}^* P_R \right) \, \frac{\xout{k}-\xout{r}+m_{\ell_\gamma}}{(k-r)^2-m_{\ell_\gamma}^2} K_{\gamma \gamma} \gamma_5  \frac{\xout{k} +\xout{p} + m_{\ell_\gamma}}{(k+p)^2-m_{\ell_\gamma}^2} \nonumber \\
&\, \left( G_{\gamma \alpha} P_L + G_{\alpha \gamma}^* P_R \right) \, \frac{1}{k^2-m_\rho^2} \, u_{\ell_\alpha} \, , \\
\M_\sigma =& \, \M_\rho \left( G \leftrightarrow C \right) \, , \\
\mathcal{M}_{X} =&  \, \bar{v}_{\ell_\beta} \sum_{j} \int \frac{d^4k}{(2 \pi)^4}\, \left( T_L^{\beta j} \, P_L  + T_R^{\beta j} \,  P_R \right) \frac{\xout{k}-\xout{r}+m_{q_j}}{(k-r)^2-m_{q_j}^2} \,
I_{j j} \, \gamma_5 \, \frac{\xout{k} +\xout{p} + m_{q_j}}{(k+p)^2-m_{q_j}^2}\,  \nonumber \\
&\left[ \left(T_L^{\alpha j}\right)^* \, P_R  + \left(T_R^{ \alpha j} \right)^*\,  P_L \right] \frac{1}{k^2 - M_{X}^2}\, u_{\ell_\alpha} \, , 
\end{align}
\begin{align}
\mathcal{M}_{Y} =&  \, \bar{v}_{\ell_\beta} \sum_{j} \int \frac{d^4k}{(2 \pi)^4}\, \gamma_\mu \left( H_L^{\beta j} \, P_L  + H_R^{\beta j} \, P_R \right) \frac{1}{k^2-M_Y^2} \left(g^{\mu \nu} - \frac{k^\mu k^\nu}{M_Y^2} \right)  \frac{\xout{k}-\xout{r}+m_{q_j}}{(k-r)^2-m_{q_j}^2} \nonumber \\
& I_{j j} \gamma_5 \, \frac{\xout{k} +\xout{p} + m_{q_j}}{(k+p)^2-m_{q_j}^2}\,\gamma_\nu  \left[ \left(H_L^{\alpha j}\right)^* \, P_L  + \left(H_R^{ \alpha j} \right)^* \, P_R \right] \, u_{\ell_\alpha} \, , \\
\M_{\sigma n} =& -2\sum_{i,j}\int \frac{d^4k}{(2 \pi)^4} \, \Tr\left[\left( B_{ij} P_R + B_{ji}^* P_L \right) \, \frac{\xout{k}+\xout{q}+m_j}{(k+q)^2-m_j^2} \, \left( A_{ij} P_R + A_{ji}^* P_L \right) \, \frac{\xout{k}+m_i}{k^2-m_i^2} \right] \nonumber \\
&\bar{v}_{\ell_\beta} \, \left( C_{ \beta \alpha} P_L + C_{ \alpha \beta}^* P_R \right) \, \frac{1}{-m_\sigma^2} u_{\ell_\alpha} \, , \\
\M_{\sigma\ell} =& -\sum_{\gamma}\int \frac{d^4k}{(2 \pi)^4} \, \Tr\left[K_{\gamma \gamma} \gamma_5 \, \frac{\xout{k}+\xout{q}+m_{\ell_\gamma}}{(k+q)^2-m_{\ell_\gamma}^2} \, C_{\gamma \gamma}^* \gamma_5\, \frac{\xout{k}+m_{\ell_\gamma}}{k^2-m_{\ell_\gamma}^2} \right] \nonumber \\
&\bar{v}_{\ell_\beta} \, \left( C_{ \beta \alpha} P_L + C_{ \alpha \beta}^* P_R \right) \, \frac{1}{-m_\sigma^2} \, u_{\ell_\alpha} \, ,\\
\M_S =& \,\sum_\gamma \bar{v}_{\ell_\beta} \int \frac{d^4k}{(2 \pi)^4} \, \left(G_{ \beta \gamma} P_L + G_{ \gamma \beta}^* P_R  \right) \frac{ \xout{k} +m_{\ell_\gamma} }{k^2 -m_{\ell_\gamma}^2} \left( C_{ \gamma \alpha} P_L + C_{ \alpha \gamma}^* P_R \right)  \nonumber \\
 &\frac{1}{\left(k-r \right)^2 -m_\rho^2} \frac{1}{\left(k+p \right)^2 -m_\sigma^2} \, \omega \, u_{\ell_\alpha} \, .
\end{align}
The amplitudes in the previous equations have been computed with the
help of {\tt Package X}~\cite{Patel:2016fam}. Since the resulting
exact analytical expressions are very involved, we have derived
approximate expressions, valid in most scenarios of interest. We now
explain the main ingredients in our calculation:
\begin{itemize}
\item A hierarchy of scales in the neutrino sector has been assumed, namely
\begin{equation} \label{eq:hier}
  m_{\rm light} \ll m_{\rm EW} \ll M_H \, .
\end{equation}
Here $m_{\rm light}$ is a low-energy scale, to be identified with the
scale of neutrino masses or with other related low mass scales,
$m_{\rm EW} \sim m_W$ is the usual electroweak scale and $M_H$ is a
high-energy scale where the heavy mediators responsible for neutrino
masses lie. This hierarchy of scales allows us to expand our results
in powers of the small mass ratios $m_{\rm light}/m_{\rm EW}$ and
$m_{\rm EW}/M_H$ and obtain approximate (but also simpler) analytical
expressions.
\item As usual, all states at a given energy scale will be assumed to
  be degenerate at order zero in $m_{\rm light}$. For instance, all
  the heavy mediators will be assumed to have a mass $M_H$, with
  possible corrections of order $m_{\rm light}$. In practice, this
  means that $m_{\rm light}/m_{\rm EW}$ and $m_{\rm EW}/M_H$ will be
  single expansion parameters and our approximations may be invalid if
  large mass splittings exist among the states that lie at a given
  energy scale.
\item In order to properly apply the mass hierarchy in
  Eq.~\eqref{eq:hier}, we split the sum over all neutral fermion mass
  eigenstates as
  \begin{equation}
    \sum_{i,j} = \sum_{i,j \sim l} + \sum_{i,j \sim h} + \sum_{i \sim l} \sum_{j \sim h}+ \sum_{j \sim l} \sum_{i \sim h} \, \label{sumsplit} ~,
  \end{equation}
  where $\sim l$ and $\sim h$ refer to the sum over light and heavy
  neutral fermions, respectively.
\item The couplings in the interaction vertices may include the
  unitary matrix $U$, see Eqs.~\eqref{eq:rotA}-\eqref{eq:rotF}. This
  must be taken into account to properly expand our results, since the
  entries of the $U$ matrix involve the mass ratios that we use as
  expansion parameters. In fact, we will make use of the identity
  \begin{equation}
    \sum_{j \sim h} U_{rj} \, m_j \, U_{kj} = \bar{M}^\dagger_{rk} -\sum_{j \sim l} U_{rj} \, m_j \, U_{kj} \, , \label{OrderEx}
  \end{equation}
  which can be readily derived from the definition of $U$. This
  expression allows us to identify up to which order in $m_{\rm
    light}$ we must expand our analytical expressions in order to be
  fully consistent. The first term, $\bar{M}^\dagger_{rk}$, is much
  larger than the neutrino mass scale, while $\sum_{j \sim l} U_{rj}
  \, m_j \, U_{kj} \sim m_\nu$ is of order one in $m_{\rm
    light}$. Therefore, if the model under consideration predicts
  $\bar{M}_{kr} \neq 0$, then the part proportional to the light
  neutrino masses does not contribute at leading order. In contrast,
  if $\bar{M}_{kr}=0$, the dominant term will be proportional to
  $m_{\rm light}$, and we must expand up to this order. In general,
  when we expand the integrals using Eq.~\eqref{sumsplit}, we find,
  mostly, that we must compare expressions of the form $\sum_{j \sim
    h} U_{rj} \, m_j \, U_{kj}$ with $\sum_{j \sim l} U_{rj} \, m_j \,
  U_{kj}$. A priori, one can think that the first one is dominant
  because $m_j$ is much larger if $j \sim h$, rather than when $j \sim
  l$. However, we do not know the shape of the $U$ matrix. In some
  models, the suppression introduced by $U$ in the heavy block
  compensates for the suppression of the light mass. Ultimately, we
  cannot compare these expressions. For this reason, we use
  Eq.~\eqref{OrderEx} intensively to transform these expressions into
  comparable ones.~\footnote{The exception to this rule takes place
  when the heavy mass appears in the denominator, $\sum_{j \sim h}
  U_{rj} \, m_j^{-1} \, U_{kj}$. In this case, it does not make sense
  to perform this transformation, as we would artificially introduce
  the dominant term into the total sum, and this will cancel out with
  the sum over the light states. Therefore, we will include it in the
  zeroth order, as it is not explicitly proportional to any power of
  the light mass. The actual order will depend on the model under
  consideration. Then, given a model, one can certainly derive the
  form of the light neutrino mass matrix, and therefore, upon
  substitution, determine the order of this term.}
\item We have simplified the results and written them using matrix
  notation. In this process, sums over repeated indices have been
  identified as matrix products whenever possible.
\item The new scalar and vector states will be assumed to be much heavier than
  $m_{\rm light}$ and all charged leptons. In the case of leptoquarks, we impose that their masses are significantly above the electroweak scale.
\end{itemize}

Before we present our results it proves convenient to introduce some
matrices that will allow us to write them in a more compact way:
\begin{align}
\Gamma_{spj}^{n,m,t} &\equiv \sum_{k,r}\bar{A}_{kr} \left[\left( \bar{M}^\dagger \right)^n \left( \bar{M} \bar{M}^\dagger \right)^m \right]_{pk }U_{rj} m_j^tU_{sj}^* \, , \label{eq:mat1} \\
\tilde{\Gamma}_{spj}^{n,m,t} &\equiv \sum_{k,r}\bar{A}_{kr} \left[\left( \bar{M}^\dagger \right)^n \left( \bar{M} \bar{M}^\dagger \right)^m \right]_{sr} U_{kj} m_j^t U_{pj}^* \, , \label{eq:mat2} \\
\Delta_{spj}^{n,m,t} &\equiv \sum_{k,r}\bar{A}_{kr} \left[\left( \bar{M}^\dagger \right)^n \left( \bar{M} \bar{M}^\dagger \right)^m \right]_{pk}U_{rj} m_j^t U_{sj} \, , \label{eq:mat3} \\ 
\tilde{\Delta}_{spj}^{n,m,t} &\equiv \sum_{k,r}\bar{A}_{kr} \left[\left( \bar{M}^\dagger \right)^n \left( \bar{M} \bar{M}^\dagger \right)^m \right]_{sr}U_{kj} m_j^t U_{pj} \, , \label{eq:mat4} \\ 
\gamma_{sp} &\equiv  \sum_{k,r} \sum_{i,j \sim l}\bar{A}_{kr}^* U_{kj}^* U_{sj} U_{ri}^* U_{pi} \, . \label{eq:mat5}
\end{align}

We are now in position to present our results. We will do so
explicitly splitting them into different orders of $m_{\rm
  light}$. The leading order contributions, as well as the divergent
pieces (when present), will be provided here. However, one should keep
in mind that this splitting in orders of $m_{\rm light}$ is performed
based on the explicit appearance of $m_{\rm light}$ in the resulting
analytical expressions. As explained above, a term that is a priori of
order zero in $m_{\rm light}$ might actually be of order one if the
specific shape of the $U$ matrix introduces one power of the low
scale. Therefore, the leading order contributions given below might be
\textit{polluted} with higher order terms, which can in general be
safely neglected. This is no longer true if the order zero
contributions vanish or turn out to be of order one due to
cancellations. In this case, a consistent calculation requires the
consideration of all order one contributions. For this reason, the
expressions for higher order terms are given in
Appendix~\ref{sec:app1}.

First of all, the $Z$ boson contribution can be written as
\begin{equation}
\M_Z = - \frac{1}{8 \pi^2} \, \bar{v}_{\ell_\alpha} \gamma_5 \, \Gamma_Z^\alpha \, u_{\ell_\alpha}  \label{eq:MZ} \, ,
\end{equation}
where
\begin{align}
\Gamma_Z^\alpha &= \,- i  \, \frac{m_{\ell_\alpha}}{v^2} \, \textup{Im}\Big\{ \left( \Gamma_Z \right)^{\rm (div)} + \left( \Gamma_Z \right)^{(0)} + \O(m_{\rm light}) \Big\} \, , \label{eq:MZ1} 
\end{align}
with $\langle H^0 \rangle = \frac{v}{\sqrt{2}}$ the usual Higgs vacuum
expectation value (VEV) that breaks the electroweak symmetry and
\begin{align}
\left( \Gamma_Z \right)^{\rm (div)} &=  \sum_{s=1}^3 \sum_{k,r}\delta_{ks} \, \bar{A}_{kr} \left[ \bar{M}^\dagger_{rk} \left (\frac{1}{\epsilon} + \log \left(\frac{\mu^2 }{M_H^2}\right)-\frac{5}{3}\right)-M_H \sum_{j\sim h} U_{kj} U_{rj}\right]  \, , \label{eq:GammaZdiv}  \\
\left( \Gamma_Z \right)^{(0)} &=  \sum_{s=1}^3 \left( \sum_{ j \sim l }\frac{\tilde{\Gamma}_{ssj}^{1,0,0}+\Gamma_{ssj}^{1,0,0}}{6} -\sum_{j\sim h}\frac{\tilde{\Delta}_{ssj}^{0,1,-1}+\Delta_{ssj}^{0,1,-1}}{3} \right) \, . \label{eq:GammaZ0}
\end{align}
The term $\left( \Gamma_Z \right)^{\rm (div)}$ contains the
dimensional regularization divergence $\displaystyle
\frac{1}{\epsilon}$. It will only be relevant in models leading to
$\delta_{ks} \, \bar{A}_{kr} \neq 0$. Since the index $s=1,2,3$ runs
over gauge non-singlet states, this will happen in models in which the
majoron couples to non-singlet fermion representations and in this
case one expects a majoron coupling to charged leptons already at
tree-level. We have also included a finite piece in $\left( \Gamma_Z
\right)^{\rm (div)}$, the last term in Eq.~\eqref{eq:GammaZdiv},
simply because it is also proportional to $\delta_{ks} \,
\bar{A}_{kr}$ and vanishes whenever the divergent piece
does. Furthermore, the rest of the contributions can be generally
written as
\begin{equation}
\M_B = \frac{1}{8 \pi^2} \, \bar{v}_{\ell_\beta} \left( L_B^{\beta \alpha} \, P_L + R_B^{\beta \alpha} \, P_R \right) u_{\ell_\alpha}  \label{eq:MB} \, ,
\end{equation}
with $B= \left\{ W, \eta, \rho, \sigma, X, Y , \sigma n, \sigma\ell, S \right\}$. The $W$-boson
contributions are given by
\begin{align}
  L_W^{\beta \alpha} &= \frac{2 \, m_{\ell_\beta}}{v^2} \, \left\{ \left( L_W^{\beta \alpha} \right)^{(0)} + \O(m_{\rm light}) \right\}  \, , \label{eq:MW}
\end{align}
with
\begin{align}
  \left( L_W^{\beta \alpha} \right)^{(0)} &= \sum_{j \sim l}\left( \frac{\Gamma_{\alpha \beta j}^{1,0,0 \, *}}{12}+ \frac{2}{3} \Gamma_{\beta \alpha j}^{1,0,0} \right) - \sum_{j\sim h} \left( \frac{\tilde{\Delta}_{\alpha \beta j}^{0,1,-1 \, *}}{6} +\frac{7}{12}\tilde{\Delta}_{\beta \alpha j}^{0,1,-1} \right) \, .  \label{eq:GammaW0}
\end{align}
The charged scalar contribution is given by
\begin{align}
  L_\eta^{\beta \alpha} = & \, m_{\ell_\alpha} \left[ \bar{D}_L^{\beta p}\left(\bar{D}_L^{\alpha s}\right)^* \left(L_\eta^{LL}\right)^*_{sp} - \left(\bar{D}_L^{\alpha p}\right)^* \bar{D}_L^{\beta s} \left(\widetilde L_\eta^{LL}\right)_{sp} \right] \nonumber \\
  & + m_{\ell_\beta} \left[ \bar{D}_R^{\beta p}\left(\bar{D}_R^{\alpha s}\right)^* \left(L_\eta^{RR}\right)^*_{sp} - \left(\bar{D}_R^{\alpha p}\right)^* \bar{D}_R^{\beta s} \left(\widetilde L_\eta^{RR}\right)_{sp} \right] \label{eq:M+-} \\
  & + \bar{D}_L^{\beta p} \left(\bar{D}_R^{\alpha s}\right)^*\left(L_\eta^{RL}\right)^*_{sp} \, , \nonumber \label{eq:Leta}
\end{align}
where
\begin{align}
  L_\eta^{LL} =& \left(L_\eta^{LL}\right)^{(0)} + \O(m_{\rm light}) \, , \\
  L_\eta^{RR} =& \left(L_\eta^{RR}\right)^{(0)} + \O(m_{\rm light}) \, , \\
  L_\eta^{RL} =& \left(L_\eta^{RL}\right)^{(0)} + \O(m_{\rm light}) \, , \\
  \widetilde L_\eta^{LL} =& L_\eta^{LL} \, \left( f_{\left(1, \, 2,\, 5, \, 6, \,7, \, 8 \right) } \leftrightarrow f_{\left(3, \, 4,\, 9, \, 10, \,15, \, 16 \right) },\: f_{13} \leftrightarrow F_{1, -3}, \: f_{14} \leftrightarrow F_{2, -4} \right) \, , \\
  \widetilde L_\eta^{RR} =& L_\eta^{RR} \, \left( f_{\left(1, \, 2,\, 5, \, 6, \,7, \, 8 \right) } \leftrightarrow f_{\left(3, \, 4,\, 9, \, 10, \,15, \, 16 \right) },\: f_{13} \leftrightarrow F_{1, -3}, \: f_{14} \leftrightarrow F_{2, -4} \right) \label{eq:LRRtilde} \, .
\end{align}
Here we have introduced the $f$ and $F$ loop functions. Their explicit
expressions can be found in Appendix~\ref{sec:app2}. We note that in
the $Z$ and $W$ bosons contributions given above, no loop functions
appeared. This is because $m_Z \sim m_W \sim m_{\rm EW}$ and we
already expanded our results in powers of $m_{\rm EW}/M_H$. However,
in contributions involving $\eta$, $\rho$ and $\sigma$ a new mass
scale appears, namely the mass of the new heavy scalar. Then, a
complete expansion cannot be made without introducing further
assumptions on their mass scale. As already explained above, our
calculation assumes $m_\eta, m_\rho, m_\sigma \gg m_{\rm light}$ but,
for the sake of generality, we prefer to stay agnostic about how these
scalar masses compare to $M_H$. Furthermore, the notation
$f_{\left(i_1 \, , ..., \, i_n\right)} \leftrightarrow f_{\left(j_1 \,
  , ..., \, j_n\right)}$ means that the replacement $f_{i_1}
\leftrightarrow f_{j_1}$, ..., $f_{i_n} \leftrightarrow f_{j_n}$ is to
be applied. We found
\begin{align}
  \left(L_\eta^{LL}\right)^{(0)} =& \left(L_\eta^{RR}\right)^{(0)} \left(\Gamma \leftrightarrow \tilde{\Gamma},\Delta \leftrightarrow \tilde{\Delta} \right) \, , \\
  \left(L_\eta^{RR}\right)^{(0)}_{sp} =& f_7 \sum_{j\sim h}\Delta^{0,1,-1}_{spj}+\sum_j \left\{ f_8 \, \Delta^{0,1,1}_{spj}-F_{5,7}  \,\tilde{\Gamma}^{1,0,0}_{spj} -F_{6,8} \, \tilde{\Gamma}^{1,1,0}_{spj} \right\} \nonumber \\
  +& \sum_{j\sim l} \Big\{ F_{5,7,-1} \, \tilde{\Gamma}^{1,0,0}_{spj}+F_{6,8,-2} \, \tilde{\Gamma}^{1,1,0}_{spj}\Big\}  \, , \label{eq:LRR} \\
  \left( L_\eta^{RL} \right)^{(0)}_{sp} =&\gamma_{sp}^* \frac{ m_{\ell_\beta}^2 + m_{\ell_\alpha}^2 }{2 m_\eta^2} -\gamma_{sp} \frac{ m_{\ell_\beta} m_{\ell_\alpha} }{2 m_\eta^2}+ f_{11} \sum_{j\sim h}\Delta_{s p j}^{1,1,-1} \nonumber \\
  +& \sum_j\left[ -f_{10} \left(\tilde{\Gamma}_{psj}^{0,2,0}\right)^* - f_9 \left(\tilde{\Gamma}_{psj}^{0,1,0}\right)^*-F_{3,11} \, \tilde{\Delta}_{s p j}^{1,0,1} +f_{12} \, \Delta_{s p j}^{1,1,1} -F_{4,12} \, \tilde{\Delta}_{s p j}^{1,1,1} \right] \nonumber \\
  +& \sum_{j\sim l}\left[F_{9,-1}\left(\tilde{\Gamma}_{psj}^{0,1,0}\right)^*+ F_{10,-2} \left(\tilde{\Gamma}_{psj}^{0,2,0}\right)^* +F_{3,-9}\left(\Gamma_{psj}^{0,1,0}\right)^*+F_{4,-10}\left(\Gamma_{psj}^{0,2,0}\right)^*\right] \label{eq:LRL} \, .
\end{align}
Let us note that we have included a term proportional not to the mass
of neutrinos but to the square of the mass of charged leptons in the
zeroth order expressions. Although this term is not explicitly
proportional to $m_{\rm light}$, we expect it to be subdominant,
simply because $m_\ell \ll M_H$. However, in the event that both
remaining terms of order zero and order one were to cancel out, then
this term would dominate over a possible order two, as $m_{\rm light}
\ll m_\ell$. We move on to the contribution induced by the neutral
scalar $\rho$, which can be written as
\begin{align} \label{eq:Lrho}
  L_\rho^{\beta \alpha} &= \frac{1}{2 m_{\rho}^2} \left[ \left( L_\rho^{\beta \alpha} \right)^{\rm (div)} +  \left( L_\rho^{\beta \alpha} \right)^{(0)} \right] \, ,
\end{align}
with
\begin{align}
 \left( L_\rho^{\beta \alpha} \right)^{\rm (div)} &= -\frac{1}{2}\left[\left(m_{\ell_\beta}^2+m_{\ell_\alpha}^2\right) +2 m_{\rho}^2 \left(1 +\frac{1}{\epsilon} +\log \frac{\mu^2}{m_{\rho}^2} \right) \right] \left(G K  G\right)_{\beta \alpha} \, , \\
 \left( L_\rho^{\beta \alpha} \right)^{(0)} &= m_{\ell_\beta}\left[ \left(G^\dagger \left( \id_3 + \log \frac{M_\ell^2}{m_{\rho}^2} \right) M_\ell K G \right)_{\beta \alpha} - \hc \right]  - \left(G \log \frac{M_\ell^2}{m_{\rho}^2}  M_\ell^2 K  G \right)_{\beta \alpha}  \nonumber \\
 & +\frac{ m_{\ell_\alpha} m_{\ell_\beta} }{2} \left(G^\dagger K  G^\dagger \right)_{\beta \alpha} \, .
\end{align}
where $\id_n$ is the $n \times n$ identity matrix. Again, we have
included a finite piece in $\left( L_\rho^{\beta \alpha} \right)^{\rm
  (div)}$ because it vanishes whenever the divergent piece does. A
similar triangle diagram contribution is induced by the pseudoscalar
$\sigma$, with
\begin{align} \label{eq:Lsigma}
  L_\sigma^{\beta \alpha} &= L_\rho^{\beta \alpha} \left( G \leftrightarrow C, m_\rho \leftrightarrow m_\sigma \right) \, .
\end{align}
The scalar leptoquark contribution is given by
\begin{align} \label{eq:LX}
  L_{X}^{\beta \alpha} &= \frac{1}{2 M_{X}^2} \left[ \left( L_{X} ^{\beta \alpha} \right)^{\rm (div)} +  \left( L_{X} ^{\beta \alpha} \right)^{(0)} \right] \, ,
\end{align}
with
\begin{align}
 \left( L_{X} ^{\beta \alpha} \right)^{\rm (div)} &= -\frac{1}{2}\left[\left(m_{\ell_\beta}^2+m_{\ell_\alpha}^2\right) +2 M_{X}^2 \left(1 +\frac{1}{\epsilon} +\log \frac{\mu^2}{M_{X}^2} \right) \right] \left(T_L I  T_R^\dagger \right)_{\beta \alpha} \, , \\
 \left( L_{X} ^{\beta \alpha} \right)^{(0)} &= m_{\ell_\beta}\left[ \left(T_R \left( \id_3 + \log \frac{M_\ell^2}{M_{X}^2} \right) M_\ell I T_R^\dagger \right)_{\beta \alpha} - \hc \right]  - \left(T_L \log \frac{M_\ell^2}{M_{X}^2}  M_\ell^2 I  T_R^\dagger \right)_{\beta \alpha}  \nonumber \\
 & +\frac{ m_{\ell_\alpha} m_{\ell_\beta} }{2} \left(T_R I  T_L^\dagger \right)_{\beta \alpha} \, .
\end{align}
In the same way, we find for the vector leptoquark contribution
\begin{align} \label{eq:LY}
L_{Y}^{\beta \alpha}= \frac{1}{M_{Y}^2} \left[ \left(L_{Y}^{\beta \alpha} \right)^{(\text{div})}   +\left(L_{Y}^{\beta \alpha} \right)^{(0)}  \right] \, ,
\end{align}
where 
\begin{align}
  \left(L_{Y}^{\beta \alpha} \right)^{(\text{div})} =&\frac{1}{2} \Big[ M_\ell \,  H_L \, M_Q I\,H_L^\dagger + H_R \, M_Q I\,H_R^\dagger \,M_\ell +2 M_\ell H_L I H_R^\dagger M_\ell \nonumber \\
    &+2 H_R \left( M_Q^2 -3 M_{Y} \id_3 \right) I H_L^\dagger \Big]_{\beta \alpha} \left( \frac{1}{\epsilon} + \log \frac{\mu^2}{M_{Y}^2}  \right) 
\end{align}
and
\begin{align}
\left(L_{Y}^{\beta \alpha} \right)^{(0)}  = &\frac{1}{4} M_\ell H_L M_Q\left( 8 \log \frac{M_{Y}^2}{M_Q^2} -7 \, \id_3 \right) I H_L^\dagger+\frac{1}{4}H_R M_Q\left( 8 \log \frac{M_{Y}^2}{M_Q^2} -7 \, \id_3 \right) I H_R^\dagger  M_\ell \nonumber \\
& + M_\ell H_L I H_R^\dagger M_\ell + H_R \left( M_Q^2+ 4 M_Q^2 \log  \frac{M_{Y}^2}{M_Q^2}  -M_{Y}^2 \id_3 \right) I H_L^\dagger
\end{align}

Note that the expressions in Eqs.~\eqref{eq:Lrho},
\eqref{eq:Lsigma} , \eqref{eq:LX} and \eqref{eq:LY} are exactly of order zero (no expansion has been
made). This is easy to understand since there are no neutrinos in the
associated loops. Lastly, the $\sigma n$ and $\sigma \ell$
contributions are given by
\begin{align}
  L_{\sigma n}^{\beta \alpha} &= \frac{1}{m_{\sigma}^2} \, C_{\beta \alpha}  \left[ \left( L_{\sigma n} \right)^{(\text{div})}+\left( L_{\sigma n} \right)^{(0)} + \O(m_{\rm light}) \right] \, , \label{eq:Lsigman}
\end{align}
with
\begin{align}
\left( L_{\sigma n} \right)^{(\text{div})} &= \text{Re} \left\{ \sum_{j} \left[ \bar{B}_{sp} \Delta_{psj}^{1,0,1}+ \bar{B}_{sp}^* \left( \Gamma_{spj}^{0,1,0} +\tilde{\Gamma}_{spj}^{0,1,0} \right) \right] \left( 3 + 2\, \frac{1}{\epsilon} + 2\,  \log \frac{\mu^2}{M_H^2}\right)  \right\} \, , \label{eq:Lsigman1} \\
\left( L_{\sigma n} \right)^{(0)}&= \text{Re} \Bigg\{ \sum_j 2 \bar{B}_{sp}^*\left[ \tilde{\Gamma}_{spj}^{0,1,0} - \Gamma_{spj}^{0,1,0}- \frac{1}{M_H^2}\left( \tilde{\Gamma}_{spj}^{0,2,0}+ \tilde{\Gamma}_{spj}^{0,1,2}\right) \right] -\frac{2}{M_H^2} \bar{B}_{sp} \tilde{\Delta}_{psj}^{1,1,1} \nonumber \\
&-\sum_{j\sim l} [ \bar{B}_{sp}^* \left( \frac{2}{M_H^2}\Gamma_{spj}^{0,2,0} -3 \Gamma_{spj}^{0,1,0}+ \tilde{\Gamma}_{spj}^{0,1,0} \right) - \sum_{j\sim h} \bar{B}_{sp} \Delta_{psj}^{1,1,-1} \Bigg\} \, , \label{eq:Lsigman2}
\end{align}
and
\begin{align}
  L_{\sigma\ell}^{\beta \alpha} &= - \frac{1}{m_\sigma^2} \, C_{\beta \alpha} \, \Tr\left[ \left(  L_{\sigma\ell} \right)^{\rm (div)} +\left(  L_{\sigma\ell} \right)^{(0)} \right] \, ,
\end{align}
with
\begin{align}
\left( L_{\sigma\ell} \right)^{\rm (div)}  &= K M_\ell^2 C^\dagger \left( I + \frac{1}{\epsilon} \right) \, \\
\left( L_{\sigma\ell} \right)^{(0)} &=K M_\ell^2 \log \frac{\mu^2}{M_\ell^2} C^\dagger \, .
 \end{align}
The cubic scalar interaction contribution leads to
\begin{align}
L_S^{\beta \alpha} = \frac{\omega}{4 \left( m_\rho^2 - m_\sigma^2 \right)^2}\left( L_{\sigma\rho}^{\beta \alpha} \right)^{(0)} \, ,
 \end{align}
with
\begin{align}
\left( L_S^{\beta \alpha} \right)^{(0)}  &= \left( G \, C^\dagger \, M_\ell \right)_{\beta \alpha} \left( -m_\rho^2 + m_\sigma^2 +m_\rho^2 \log \frac{m_\rho^2}{m_\sigma^2} \right) - \left( M_\ell G^\dagger \, C  \right)_{\beta \alpha} \left( -m_\rho^2 + m_\sigma^2 +m_\sigma^2 \log \frac{m_\rho^2}{m_\sigma^2} \right) \nonumber \\
&-2 \left( G \, M_\ell \, C \right)\left( m_\rho^2 - m_\sigma^2 \right) \log \frac{m_\rho^2}{m_\sigma^2}  \, .
 \end{align}
Again, the last three expressions are exactly of order zero in $m_{\rm
  light}$ since there are no neutrinos in the loop. We also note that
only the real parts contribute in Eqs.~\eqref{eq:Lsigman1} and
\eqref{eq:Lsigman2}. To conclude, the right chirality couplings are
given by
\begin{equation}
R_B^{\beta \alpha} = L_B^{\alpha \beta \, \ast} \, .
\end{equation}
Finally, the majoron coupling to a pair of charged leptons, defined in
Eq.~\eqref{eq:llJ}, can be written in terms of the results presented
in this Section as
\begin{equation} \label{eq:Scoup}
S^{\beta \alpha} = \frac{1}{8 \pi^2}\left( \delta^{\beta \alpha} \, \Gamma_Z^\alpha + \sum_B L_B^{\beta \alpha} \right) \, .
\end{equation}

Some comments are in order. First of all, we emphasize that these
expressions are very long because they are meant to be completely
general. In specific models, most of the terms will simply vanish or
take very simple forms, as we will show explicitly in
Sec.~\ref{sec:examples}. Secondly, while most contributions are
perfectly finite, some include divergent pieces. As already explained,
these only appear in contributions that would necessarily lead to the
existence of a majoron coupling to a pair of charged leptons already
at tree-level. In this case they are expected to induce just small
(finite) corrections, but we included them for the sake of
completeness.

\section{Application to specific models}
\label{sec:examples}

In this Section we show several example models and illustrate how our
general results for the majoron coupling to charged leptons apply to
these specific cases. We consider several neutrino mass models, some
of them very well known. Instead of breaking lepton number explicitly,
as usually done, we promote it to a global conserved \U1L symmetry
that gets spontaneously broken by the vacuum expectation value (VEV)
of a scalar $\chi$. This implies the presence of a majoron in the
particle spectrum of the theory.

\subsection{Type-I seesaw}
\label{subsec:typeI}

{
\renewcommand{\arraystretch}{1.4}
\begin{table}[t!]
\centering
{\setlength{\tabcolsep}{0.5em}
\begin{tabular}{c c c c c c c}
\toprule  
field & spin & generations & $\mathrm{SU(3)}_c$ & $\mathrm{SU(2)}_L$ & $\mathrm{U(1)}_Y$ & \U1L \\
\midrule   
$N$ & $\frac{1}{2}$ & 3 & \one & \one & 0 & 1 \\
\midrule
$\chi$ & 0 & 1 & \one & \one & 0 & -2 \\
\bottomrule
\end{tabular}
}
\caption{New particles in the type-I seesaw with spontaneous lepton number violation.}
\label{tab:typeI}
\end{table}
}

The type-I seesaw with spontaneous lepton number
violation~\cite{Chikashige:1980ui} extends the SM particle content
with 3 singlet fermions $N$ and a scalar singlet $\chi$, all charged
under the global \U1L as shown in Tab.~\ref{tab:typeI}. The Yukawa
terms relevant for our discussion are
\begin{equation}
    -\mathcal{L} = y \, \bar{L} \tilde{H} N + \frac{1}{2} \, \lambda \, \chi \, \bar{N}^c N + \hc \, . 
\end{equation}
Here $H$ is the SM Higgs doublet, with $\tilde{H} = i \sigma_2 H^*$,
$y$ a general $3 \times 3$ matrix and $\lambda$ a symmetric $3 \times
3$ matrix. We will work in the basis in which $\lambda$ is a diagonal
matrix with real entries. The $\chi$ singlet can be decomposed as
\begin{equation} \label{eq:sigma}
  \chi = \frac{1}{\sqrt{2}} \left( v_\chi + \rho + i J \right) \, ,
\end{equation}
where $\vev{\chi} = \frac{v_\chi}{\sqrt{2}}$ is the $\chi$ VEV
responsible for the breaking of \U1L, $\rho$ is a massive CP-even
scalar and $J$ is the massless majoron, the Goldstone boson associated
to the spontaneous violation of lepton number. After symmetry
breaking, a Dirac mass term, $M_D$, that mixes the left-handed
neutrinos in the lepton doublets with the singlet neutrinos, as well
as Majorana mass term for the singlet neutrinos, $M_R$, are
induced. They are given by
\begin{equation} \label{eq:MDMR}
  M_D = y \, \frac{v}{\sqrt{2}} \, , \quad M_R = \lambda \, \frac{v_\chi}{\sqrt{2}} \, .
\end{equation}
In the basis $\left\{ \nu_L^c , N \right\}$, the resulting $6 \times
6$ Majorana mass matrix for the neutral fermions can be written as
\begin{equation} \label{eq:typeIbarM}
   \bar{M} = \begin{pmatrix}
    0 & M_D \\
    M_D^T & M_R
    \end{pmatrix} \, .
\end{equation}
If one assumes the hierarchy $M_D \ll M_R$ (equivalent to $m_{\rm EW}
\ll M_H$ in Eq.~\eqref{eq:hier}), the light neutrinos mass matrix is
given by $m_\nu = -M_D \, M_R^{-1} \, M_D^T$, hence leading to
naturally suppressed masses. This is the usual type-I seesaw
mechanism~\cite{Minkowski:1977sc,Yanagida:1979as,Mohapatra:1979ia,Gell-Mann:1979vob,Schechter:1980gr}.

Since $\rho$ does not couple to the charged leptons at tree-level, the
only diagrams that induce a majoron coupling to charged leptons are
those with gauge bosons, shown in Fig.~\ref{fig:Wdiag} and
\ref{fig:Zdiag}. Therefore, we only need to adapt the general results
for $\Gamma_Z$ and $L_W$ to the type-I seesaw. We first identify the
general couplings of Eqs.~\eqref{eq:rotA}-\eqref{eq:rotF} that are
involved in these contributions. In this model, the majoron coupling
to a pair a neutral fermions is given in the gauge basis by
\begin{equation}
  \bar{A}_{kr} = \left\{ \begin{array}{cl}
    \displaystyle  \frac{i}{2 v_\chi} \, \bar{M}_{kr} \, , & \text{if} \,\, k, r = 4,5,6 \\
     & \\
    \displaystyle 0 \, , & \text{otherwise} \end{array} \right.
\end{equation}
The $A$, $E$ and $F$ couplings in the mass basis are then given by
Eqs.~\eqref{eq:rotA}, \eqref{eq:rotE} and \eqref{eq:rotF}, although
they are not necessary for our computation. First, we note that
$\left( \Gamma_Z \right)^{\rm (div)}$ vanishes exactly since
$\delta_{ks} \bar{A}_{kr} = 0$ for $s=1,2,3$. Then, we just need to
compute $\sum_{ j \sim l} \tilde{\Gamma }_{ \beta \alpha j}^{1,0,0}$,
$\sum_{ j \sim l} \Gamma_{\beta \alpha j}^{1,0,0} $, $\sum_{ j \sim h}
\Delta_{\beta \alpha j}^{0,1,-1}$ and $\sum_{ j \sim h}
\tilde{\Delta}_{ \beta \alpha j}^{0,1,-1}$. In order to do that we
need to know the form of the $U$ matrix. In the type-I seesaw this
unitary matrix can be written as~\cite{Grimus:2000vj}
\begin{equation} \label{eq:Umatrix}
    U=\begin{pmatrix}
    U_l & 0 \\
    0 & U_h
    \end{pmatrix} \, \begin{pmatrix}
    \sqrt{\id_3 - P P^\dagger} & P \\
    -P^\dagger & \sqrt{\id_3 - P^\dagger P}
    \end{pmatrix} \equiv U_2 \, U_1 \, .
\end{equation}
Here $U_1$ brings the neutral fermions mass matrix into a
block-diagonal form, while $U_2$ finally diagonalizes, independently,
the light and heavy sectors of the matrix. $U_l$, $U_h$ and $P$ are $3
\times 3$ matrices. The matrix $P$ depends on the matrices $M_D$ and
$M_R$ in a non-trivial way. However, one can expand $P$ in inverse
powers of the large $M_R$ scale as
\begin{equation} \label{eq:Pmatrix}
    P=\sum_{i=1}^\infty P_i \, ,
\end{equation}
with $P_i \propto M_R^{-i}$. At leading order in $M_R^{-1}$, one finds
\begin{align}
  \sqrt{\id_3 - P P^\dagger} = \sqrt{\id_3 - P^\dagger P} = \id_3 + \O(M_R^{-2}) \, ,
\end{align}
and
\begin{align}
  P = P_1 + \O(M_R^{-3}) = M_D^* \, M_R^{-1} + \O(M_R^{-3}) \, .
\end{align}
With these expressions at hand it is straightforward to expand the
sums in Eqs.~\eqref{eq:mat1}-\eqref{eq:mat4} and find
\begin{align}
  \sum_{ j \sim l } \Gamma_{\beta \alpha j}^{1,0,0 } &= \frac{i}{2 v_\chi} \sum_{k,r=4}^6 \sum_{j=1}^3 \bar{M}_{kr} \bar{M}^\dagger_{\alpha k} U_{rj} U_{\beta j}^* \nonumber \\
  &\simeq \frac{i}{2 v_\chi} \left(- \id_3 \, M_D M_R^{-1} M_R M_D^\dagger \right)_{\beta \alpha} = -\frac{i}{2 v_\chi} \left( M_D M_D^\dagger \right)_{\beta \alpha} \, , \\
  \sum_{ j \sim l } \tilde{\Gamma}_{\beta \alpha j}^{1,0,0 } &= \frac{i}{2 v_\chi} \sum_{k,r=4}^6 \sum_{j=1}^3 \bar{M}_{kr} \bar{M}^\dagger_{\beta r} U_{kj} U_{\alpha j}^* \nonumber \\
  &\simeq \frac{i}{2 v_\chi} \left(- M_D^* M_R M_R^{-1}  M_D^T \, \id_3 \right)_{\beta \alpha} = -\frac{i}{2 v_\chi} \left(  M_D M_D^\dagger \right)_{ \alpha \beta} \, , \\
  \sum_{ j \sim h} \Delta_{ \beta \alpha  j}^{0,1,-1} &= \frac{i}{2 v_\chi} \sum_{k,r=4}^6 \sum_{j=4}^6 \bar{M}_{kr} \left(\bar{M} \bar{M}^\dagger \right)_{\alpha k} U_{rj} m_j^{-1} U_{\beta j} \nonumber \\
  &\simeq \frac{i}{2 v_\chi}  \left( M_D^* M_R^{-1} M_R^{-1} M_R M_R M_D^T \right)_{\beta \alpha} = \frac{i}{2 v_\chi} \left( M_D M_D^\dagger \right)_{\alpha \beta } \, , \\
  \sum_{ j \sim h} \tilde{\Delta}_{ \beta \alpha j}^{0,1,-1} &= \frac{i}{2 v_\chi} \sum_{k,r=4}^6 \sum_{j=4}^6 \bar{M}_{kr} \left( \bar{M} \bar{M}^\dagger \right)_{\beta r} U_{kj} m_j^{-1} U_{\alpha j} \nonumber \\
  &\simeq \frac{i}{2 v_\chi} \left( M_D M_R M_R M_R^{-1} M_R^{-1} M_D^\dagger \right)_{\beta \alpha} = \frac{i}{2 v_\chi} \left( M_D M_D^\dagger \right)_{\beta \alpha} \, .
\end{align}
At this point we just need to combine all these results into a final
expression for the majoron coupling to charged leptons. For the $Z$
boson contribution we make use of Eq.~\eqref{eq:GammaZ0} to obtain
\begin{equation}
  \left( \Gamma_Z \right)^{(0)} = - \frac{i}{2 v_\chi} \, \textup{Tr}(M_D M_D^\dagger) \, ,
\end{equation}
whereas for the $W$ boson contribution, using Eq.~\eqref{eq:GammaW0},
we find
\begin{equation}
  \left( L_W^{\beta \alpha} \right)^{(0)} = - \frac{i}{2 v_\chi} \, \left( M_D M_D^\dagger \right)_{\beta \alpha} \, .
\end{equation}
Finally, replacing these contributions into Eqs.~\eqref{eq:MZ1} and
\eqref{eq:MW}, and using the generic Lagrangian in Eqs.~\eqref{eq:llJ}
and the expression for the coupling $S$ in Eq.~\eqref{eq:Scoup}, we
recover the known formula for the majoron coupling to charged leptons
in the type-I seesaw~\cite{Pilaftsis:1993af,Garcia-Cely:2017oco,Heeck:2019guh},
\begin{equation}
\mathcal{L}_{\ell\ell J} =\frac{iJ}{16\pi^2 v^2 v_\chi} \bar{\ell}\left[M_\ell ~ \textup{Tr}(M_D M_D^\dagger)\gamma_5 +2 M_\ell M_D M_D^\dagger P_L-2  M_D M_D^\dagger M_\ell P_R\right]\ell \, .
\end{equation}
The \textit{natural} size of these couplings can be readily estimated
from this expression. The coupling $S$ in Eq.~\eqref{eq:Scoup} is
estimated as
\begin{equation} \label{eq:StypeI}
  S_{\rm Type-I} \sim \frac{M_\ell \, M_D^2}{16 \pi^2 \, v^2 \, v_\chi} \sim \frac{M_\ell \, m_\nu}{16 \pi^2 \, v^2} \, ,
\end{equation}
where we have used that $m_\nu \sim M_D^2 / M_R$ and $M_R \sim v_\chi$
in this model. The resulting coupling is strongly suppressed, not only
by the loop factor, but also by charged lepton and neutrino
masses. For instance, with $M_\ell \sim m_\mu$ and $m_\nu \sim 0.1$
eV, one finds $S_{\rm Type-I} \sim 10^{-18}$. This generic result
clearly respects the bounds in Sec.~\ref{sec:coup}
(Eqs.~\eqref{eq:See}, \eqref{eq:Smumu}, \eqref{eq:mulim} and
\eqref{eq:taulim}). In fact, using Eq.~\eqref{eq:gamma} one finds
branching ratios as low as BR$(\mu \to e \, J) \sim 10^{-20}$, well
below any foreseen experimental search.

In summary, in the type-I seesaw one generally expects tiny majoron
couplings to the charged leptons. The only way out of this conclusion
is to take advantage of the fact that neutrino masses and the majoron
coupling to charged leptons actually depend on a different product of
matrices ($M_D M_R^{-1} M_D^T$ versus $M_D M_D^\dagger /
v_\chi$). Therefore, one can in principle evade the generic
expectation of Eq.~\eqref{eq:StypeI} by adopting specific textures for
the $y$ Yukawa matrix.

\subsection{Inverse seesaw}
\label{subsec:inverse}

{
\renewcommand{\arraystretch}{1.4}
\begin{table}[tb]
\centering
{\setlength{\tabcolsep}{0.5em}
\begin{tabular}{c c c c c c c}
\toprule  
field & spin & generations & $\mathrm{SU(3)}_c$ & $\mathrm{SU(2)}_L$ & $\mathrm{U(1)}_Y$ & \U1L \\
\midrule   
$N$ & $\frac{1}{2}$ & 3 & \one & \one & 0 & -1 \\
$S$ & $\frac{1}{2}$ & 3 & \one & \one & 0 & 1 \\
\midrule
$\chi$ & 0 & 1 & \one & \one & 0 & -2 \\
\bottomrule
\end{tabular}
}
\caption{New particles in the inverse seesaw with spontaneous lepton number violation.}
\label{tab:ISS}
\end{table}
}

The inverse seesaw with spontaneous lepton number violation~\cite{CentellesChulia:2020dfh} introduces
the new fields in Tab.~\ref{tab:ISS}. In this case, in addition to the
usual 3 $N$ fermion singlets, 3 $S$ fermion singlets are included too,
with opposite lepton number. The relevant Yukawa Lagrangian terms are
given by
\begin{equation}
-\mathcal{L} = y \, \bar{L} \tilde{H} N + M_R \, \bar{N}^c S + \frac{1}{2} \, \lambda' \, \chi^* \bar{N}^c N + \frac{1}{2} \, \lambda \, \chi \bar{S}^c S + \hc \, .
\end{equation}
We note the presence of two Yukawa terms involving $\chi$, with
couplings $\lambda'$ and $\lambda$, two symmetric $3 \times 3$
matrices. Moreover, $M_R$ is a general $3 \times 3$ matrix with
dimensions of mass. However, we will work in the basis in which $M_R$
is a diagonal matrix with real entries. This choice is particularly
convenient for our computation, since $M_R$ will be taken below as
expansion parameter. Again, the singlet $\chi$ can be decomposed as in
Eq.~\eqref{eq:sigma}, including a massless majoron, $J$. Furthermore,
symmetry breaking leads to a Dirac mass term, $M_D$, as in the type-I
seesaw, as well as to Majorana mass terms for the singlet neutrinoss,
$\mu'$ and $\mu$, given by
\begin{equation} 
  M_D = y \, \frac{v}{\sqrt{2}} \, , \quad \mu' = \lambda' \, \frac{v_\chi}{\sqrt{2}} \, , \quad \mu = \lambda \, \frac{v_\chi}{\sqrt{2}} \, .
\end{equation}
In the basis $\left\{ \nu_L^c , N , S^c \right\}$, the $9 \times 9$ Majorana
mass matrix for the neutral fermions can be written as
\begin{equation} \label{eq:inversebarM}
   \bar{M} = \begin{pmatrix}
    0 & M_D & 0 \\
    M_D^T & \mu' & M_R \\
    0 & M_R & \mu
    \end{pmatrix} \equiv \begin{pmatrix}
    0 & m_D \\
    m_D^T & Q
    \end{pmatrix} \, ,
\end{equation}
with
\begin{equation} \label{eq:Pmatrix}
 Q=
 \begin{pmatrix}
 \mu' & M_R \\
 M_R & \mu
 \end{pmatrix} \, ,
\end{equation}
and 
\begin{equation} \label{eq:mDmatrix}
  m_D = \begin{pmatrix}
  M_D & 0 \\
 \end{pmatrix} \, .
\end{equation}
If one assumes the hierarchy $\mu,\mu' \ll M_D \ll M_R$, the light
neutrinos mass matrix is given by $m_\nu = M_D M_R^{-1} \mu M_R^{-1}
M_D^T$, hence leading to naturally small neutrino masses. We note that
the $\mu'$ parameter does not contribute to neutrino masses at leading
order in the expansion.

In this model, just as in the type-I seesaw, the only diagrams
contributing to the majoron coupling to charged leptons are those with
gauge bosons, shown in Fig.~\ref{fig:Wdiag} and \ref{fig:Zdiag}. Then,
we only need to compute $\Gamma_Z$ and $L_W$ at leading order. In
order to do that we must identify the general couplings of
Eqs.~\eqref{eq:rotA}-\eqref{eq:rotF} that participate in these
contributions. In the gauge basis, the majoron coupling to a pair a
neutral fermions is given by
\begin{equation}
  \bar{A}_{kr} = \left\{ \begin{array}{cl}
    \displaystyle -\frac{i}{2 v_\chi} \, \bar{M}_{kr} \, , & \text{if} \,\, k, r = 4,5,6 \\
    & \\
    \displaystyle \frac{i}{2 v_\chi} \, \bar{M}_{kr} \, , & \text{if} \,\, k, r = 7,8,9 \\
    & \\
    \displaystyle 0 \, , & \text{otherwise}  \end{array} \right.
\end{equation}
The $A$, $E$ and $F$ couplings in the mass basis can be readily
obtained with the help of Eqs.~\eqref{eq:rotA}, \eqref{eq:rotE} and
\eqref{eq:rotF}. Again, the divergent piece $\left( \Gamma_Z
\right)^{\rm (div)}$ vanishes exactly due to $\delta_{ks} \bar{A}_{kr}
= 0$ for $s=1,2,3$. Therefore, we must compute $\sum_{j \sim l}
\tilde{\Gamma}_{\beta \alpha j}^{1,0,0}$, $\sum_{j \sim l}
\Gamma_{\beta \alpha j}^{1,0,0}$, $\sum_{j \sim h} \Delta_{\beta
  \alpha j}^{0,1,-1}$ and $\sum_{j \sim h} \tilde{\Delta}_{ \beta
  \alpha j}^{0,1,-1}$. In order to do that we need the form of the $U$
matrix, which in this model is a $9 \times 9$ unitary matrix. When the
mass matrix $\bar M$ in Eq.~\eqref{eq:inversebarM} is written in terms
of the matrices $Q$ and $m_D$ in Eqs.~\eqref{eq:Pmatrix} and
\eqref{eq:mDmatrix}, one obtains an expression that is formally
equivalent to that in the type-I seesaw, see
Eq.~\eqref{eq:typeIbarM}. Therefore, $U$ takes the same form, namely
\begin{equation}
  U=\begin{pmatrix}
    U_l & 0 \\
    0 & U_h
    \end{pmatrix} \,
  \begin{pmatrix}
    \sqrt{\id_3 - P P^\dagger} & P \\
    -P^\dagger & \sqrt{\id_6 - P^\dagger P}
    \end{pmatrix} \equiv U_2 \, U_1 \, ,
\end{equation}
with $U_l$ and $U_h$ two $3 \times 3$ and $6 \times 6$ matrices,
respectively. $P$ is a $3 \times 6$ matrix that can be expanded in
inverse powers of the large $M_R$ scale as in
Eq.~\eqref{eq:Pmatrix}. At leading order one finds
\begin{align}
  \sqrt{\id_3 - P P^\dagger} &= \id_3 + \O(M_R^{-2}) \, , \\
  \sqrt{\id_6 - P^\dagger P} &= \id_6 + \O(M_R^{-2}) \, ,
\end{align}
and
\begin{align}
  P = P_1 + \O(M_R^{-3}) = m_D^* \, \left(Q^{-1}\right)^\dagger + \O(M_R^{-3}) \, .
\end{align}
With these expressions, one can readily find
\begin{align}
  \sum_{j \sim l } \Gamma_{\beta \alpha j}^{1,0,0} &= \frac{i}{2 v_\chi}  \sum_{j =1}^3 \left( \sum_{k,r=7}^9 -\sum_{k,r=4}^6 \right) \bar{M}_{kr} \left( \bar{M}^\dagger \right)_{\alpha k} U_{rj} U_{\beta j}^* \nonumber \\
  &\simeq - \frac{i}{2 v_\chi}  \left( M_D   M_R^{-1} \mu \, M_R^{-1} \mu' M_D^\dagger \right)_{\beta \alpha} \, , \\
  \sum_{j \sim l} \tilde{\Gamma}_{\beta \alpha j}^{1,0,0} &= \frac{i}{2 v_\chi}  \sum_{j=1}^3 \left( \sum_{k,r=7}^9 -\sum_{k,r=4}^6 \right) \bar{M}_{kr} \left( \bar{M}^\dagger \right)_{\beta r } U_{kj} U_{\alpha j}^* \nonumber \\
  &\simeq - \frac{i}{2 v_\chi} \left( M_D^* \mu' M_R^{-1} \mu \, M_R^{-1}  M_D^T \right)_{\beta \alpha} \, , \\
  \sum_{j \sim h} \Delta_{\beta \alpha j}^{0,1,-1} &= \frac{i}{2 v_\chi} \sum_{j =4}^9 \left( \sum_{k,r=7}^9 -\sum_{k,r=4}^6 \right) \left( \bar{M} \bar{M}^\dagger \right)_{\alpha k} U_{rj} m_j^{-1} U_{\beta j} \nonumber \\
  &\simeq - \frac{i}{2 v_\chi}   \left( M_D^* M_R^{-1}  \mu^\dagger M_R^{-2} \mu \, M_R M_D^T +  M_D^* M_R^{-2} \mu' \mu^{' \dagger} M_D^T  + M_D^* M_R^{-2} \mu' M_R^{-1} \mu M_D^T \right)_{\beta \alpha} \, , \\
  \sum_{j \sim h} \tilde{\Delta}_{\beta \alpha j}^{0,1,-1} &= \frac{i}{2 v_\chi}  \sum_{j=4}^9 \left( \sum_{k,r=7}^9 -\sum_{k,r=4}^6 \right) \bar{M}_{kr} \left( \bar{M} \bar{M}^\dagger \right)_{\beta r} U_{kj} m_j^{-1} U_{\alpha j} \nonumber \\
  &\simeq \frac{-i}{2 v_\chi}  \left(M_D M_R \, \mu \, M_R^{-2} \mu^\dagger M_R^{-1} M_D^\dagger + M_D \mu^{' \dagger} \mu' M_R^{-2} M_D^\dagger + M_D M_R \mu M_R^{-1} \mu' M_R^{-2}M_D^\dagger \right)_{\beta \alpha}   \, .
\end{align}
Now we just need to introduce these results into the expressions for
the leading order $Z$ and $W$ boson contributions. For the $Z$ boson
contribution we use Eq.~\eqref{eq:GammaZ0} to obtain
\begin{equation}
  \left( \Gamma_Z \right)^{(0)} =  \frac{i}{6 v_\chi} \,\textup{Tr}\left( M_D \left[ \mu \mu' +2 \mu \mu^\dagger +2 \mu^{' \dagger} \mu' \right] M_R^{-2} M_D^\dagger \right) \, ,
\end{equation}
whereas for the $W$ boson contribution, using Eq.~\eqref{eq:GammaW0},
we find
\begin{equation}
  \left( L_W^{\beta \alpha} \right)^{(0)} =  \frac{i}{24 v_\chi} \, \left( M_D \left[  -\mu \mu' - \mu^{' \dagger} \mu^\dagger +5 \mu \mu^\dagger + 5 \mu' \mu^{' \dagger} \right] M_R^{-2} M_D^\dagger \right)_{\beta \alpha} \, \, .
\end{equation}
Finally, replacing these contributions into Eqs.~\eqref{eq:MZ1} and
\eqref{eq:MW}, and using the generic expressions in
Eqs.~\eqref{eq:llJ} and \eqref{eq:Scoup} we can write the coupling of
the majoron to a pair of charged leptons in the inverse seesaw as
\begin{align}
\mathcal{L}_{J\ell\ell} =- \frac{iJ}{96\pi^2 v^2 v_\chi} \bar{\ell} \, \Bigg\{&2 M_\ell \,  \textup{Tr} \left( M_D \left[ \mu \mu' +2 \mu \mu^\dagger + 2\mu^{' \dagger} \mu' \right] M_R^{-2} M_D^\dagger \right) \gamma_5 \nonumber \\
&+ M_\ell  \, M_D \left( 5 \mu \mu^\dagger + 5 \mu' \mu^{' \dagger} - \mu \mu' - \mu^{' \dagger} \mu^\dagger \right) M_R^{-2} M_D^\dagger  \, P_L  \nonumber \\
& -  M_D \left( 5 \mu \mu^\dagger + 5 \mu' \mu^{' \dagger} - \mu^{' \dagger} \mu^\dagger - \mu \mu'   \right) M_R^{-2} M_D^\dagger \, M_\ell \, P_R  \Bigg\} \,\ell \, .
\end{align}
In the case of $\mu'=0$ (or, equivalently, $\lambda'=0$), an
assumption that is often made in the literature, this expression
reduces to
\begin{align}
\mathcal{L}_{J\ell\ell} = &-\frac{iJ}{96\pi^2 v^2 v_\chi} \bar{\ell} \, \Bigg\{4 M_\ell \, \textup{Tr} \left( M_D \, \mu \mu^\dagger M_R^{-2} M_D^\dagger \right) \gamma_5 \nonumber \\
&+5 M_\ell  \, M_D \, \mu \mu^\dagger M_R^{-2} M_D^\dagger  \, P_L  -5 M_D \, \mu \mu^\dagger M_R^{-2} M_D^\dagger \, M_\ell \, P_R  \Bigg\} \,\ell \, .
\end{align}
To the best of our knowledge, this is the first time that this
coupling is computed in the context of the inverse seesaw. We can now
proceed in the same way as in the type-I seesaw and estimate its size
in the inverse seesaw. In this case, and assuming $\mu'=0$ for
simplicity, the coupling $S$ can be estimated as
\begin{equation} \label{eq:Sinverse}
  S_{\rm ISS} \sim \frac{M_\ell \, M_D^2 \, \mu^2}{96 \pi^2 \, v^2 \, v_\chi \, M_R^2} \sim \frac{M_\ell \, \mu \, m_\nu}{96 \pi^2 \, v^2 \, v_\chi} \sim \frac{M_\ell \, m_\nu}{96 \pi^2 \, v^2} \, ,
\end{equation}
where we have used that $m_\nu \sim M_D^2 \mu / M_R^2$ and $\mu \sim
v_\chi$ in this model. The resulting coupling is again very strongly
suppressed. In fact, one finds the same suppression factors as in the
type-I seesaw (see Eq.~\eqref{eq:StypeI}) and thus expects $S_{\rm
  ISS} \sim S_{\rm Type-I}$. One could in principle increase $S_{\rm
  ISS}$ by introducing a large non-vanishing $\mu'$ (or $\lambda'$),
since this coupling does not affect neutrino masses at leading
order. However, it does enter subleadingly, see for
instance~\cite{Abada:2014vea}. Therefore, $\mu'$ cannot be increased
indefinitely.

We conclude that in the version of the inverse seesaw with spontaneous
lepton number violation considered here one also expects tiny majoron
couplings to the charged leptons. Again, as in the type-I seesaw, one
can evade this conclusion by properly tuning the Yukawa matrices. This
may be used to break the generic expectation in
Eq.~\eqref{eq:Sinverse}. Furthermore, this conclusion may not hold in
alternative versions of the inverse seesaw with spontaneous lepton
number violation~\cite{Boulebnane:2017fxw,Cuesta:2021kca}.

\subsection{Scotogenic model}
\label{subsec:scoto}

{
\renewcommand{\arraystretch}{1.4}
\begin{table}[tb]
\centering
{\setlength{\tabcolsep}{0.5em}
\begin{tabular}{c c c c c c c c}
\toprule  
field & spin & generations & $\mathrm{SU(3)}_c$ & $\mathrm{SU(2)}_L$ & $\mathrm{U(1)}_Y$ & \U1L & \z2 \\
\midrule   
$N$ & $\frac{1}{2}$ & 3 & \one & \one & 0 & 1 & - \\
\midrule
$\eta$ & 0 & 1 & \one & \two & $\frac{1}{2}$ & 0 & - \\
$\chi$ & 0 & 1 & \one & \one & 0 & -2 & + \\
\bottomrule
\end{tabular}
}
\caption{New particles in the Scotogenic model with spontaneous lepton number violation.}
\label{tab:scoto}
\end{table}
}

In the Scotogenic model~\cite{Tao:1996vb,Ma:2006km} with spontaneous
lepton number violation, the particle spectrum is extended with 3
fermion singlets, $N$, a scalar doublet, $\eta$, and a scalar singlet,
$\chi$. A new \z2 parity is introduced as well, under which $N$ and
$\eta$ are odd while the rest of the fields in the model are even. The
fields and their charges under the symmetries of the model are given
in Tab.~\ref{tab:scoto}. The terms of the Yukawa Lagrangian that are
relevant for our discussion are given by
\begin{equation}
-\mathcal{L} = y \, \bar{L} \tilde{\eta} N + \frac{1}{2} \, \lambda \, \chi \bar{N}^c N + \hc \, .
\end{equation}
with $\tilde{\eta} = i \sigma_2 \eta^*$. The doublet $\eta$ is assumed
to have a vanishing VEV, $\vev{\eta} = 0$ in order to preserve the \z2
symmetry. This ensures that the lightest \z2-odd state is completely
stable. The singlet $\chi$ can be decomposed as in
Eq.~\eqref{eq:sigma}, giving rise to a massless majoron, $J$. The
breaking of \U1L induces a Majorana mass term for the singlet
fermions, $M_N$, defined as
\begin{equation}
  M_N = \lambda \, \frac{v_\chi}{\sqrt{2}} \, .
\end{equation}
Note, however, that the \z2 symmetry precludes any Dirac mass term
mixing the SM neutrinos with the fermion singlets. This in turn
implies that neutrinos remain massless at tree-level.~\footnote{In the
Scotogenic model, neutrinos get non-zero masses at the 1-loop
level. This is due to the breaking of lepton number in two units by
the $\chi$ VEV. In this work we are interested in the 1-loop coupling
of the majoron to charged leptons. Any effect on this coupling
associated to the 1-loop neutrino mass would be of higher order in
pertubation theory, hence not relevant for our discussion.} In fact,
in the basis $\left\{ \nu_L^c , N \right\}$, the tree-level $6 \times
6$ Majorana mass matrix for the neutral fermions can be written as
\begin{equation} \label{eq:ScotobarM}
   \bar{M} = \begin{pmatrix}
    0 & 0 \\
    0 & M_N
    \end{pmatrix} \, ,
\end{equation}
which trivially leads to $U = \id_6$.

In the Scotogenic model, the majoron interacts at tree-level only with
the $N$ fermion singlets, which are mass eigenstates since the
$\nu_L-N$ mixing is exactly zero. For this reason, the gauge bosons
contributions to the majoron coupling to charged leptons are absent in
this model. The only 1-loop contribution is given by the $\eta$
charged scalar, as shown in Fig.~\ref{fig:EtaRhodiag}. We just need to
compute $L_\eta$, given in Eq.~\eqref{eq:M+-}. The relevant couplings
in the gauge basis are
\begin{equation}
  \bar{A}_{kr} = \left\{ \begin{array}{cl}
    \displaystyle i \frac{M_N}{2 v_\chi} \, \delta_{kr} \, , & \text{if} \,\, k, r = 4,5,6 \\
     & \\
    \displaystyle 0 \, , & \text{otherwise} \end{array} \right.
\end{equation}
and
\begin{align}
  D_L^{\alpha i} &= 0 \, , \\
  D_R^{\alpha i} &= - \sum_{k=4}^6 y^{\alpha k} \, \delta_{ki} \, .
\end{align}
The couplings in the mass basis can be computed using
Eqs.~\eqref{eq:rotA}, \eqref{eq:rotDL} and \eqref{eq:rotDR}. Since
$D_L^{\alpha i} = 0$, we only need to compute
$\left(L_\eta^{RR}\right)^{(0)}$. This contribution involves the sums
over light states $\sum_{j \sim l} \tilde{\Gamma}^{1,0,0}_{spj}$ and
$\sum_{j \sim l} \tilde{\Gamma}^{1,1,0}_{spj}$, but these are
trivially zero, since they both involve the product $\bar A_{kr}
U_{kj} = \bar A_{kr} \delta_{kj} = 0$ for $j=1,2,3$. Therefore, we are
left with $\sum_{j \sim h} \Delta_{spj}^{0,1,-1}$, $\sum_{j}
\Delta_{spj}^{0,1,1}$, $\sum_{j} \tilde{\Gamma}_{spj}^{1,0,0}$ and
$\sum_{j} \tilde{\Gamma}_{spj}^{1,1,0}$. They are found to be
\begin{align}
\sum_{j \sim h} \Delta_{spj}^{0,1,-1} &= i \frac{M_N}{2 v_\chi} \sum_{k,r=4}^6 \delta_{kr} \left( \bar{M} \bar{M}^\dagger \right)_{pk} \delta_{rj} \, m_j^{-1} \, \delta_{sj} = i \frac{M_N^2}{2 v_\chi} \, \delta_{sp} \, , \\
\sum_{j} \Delta_{spj}^{0,1,1} &= i \frac{M_N}{2 v_\chi} \sum_{k,r=4}^6 \delta_{kr} \left( \bar{M} \bar{M}^\dagger \right)_{pk}\delta_{rj} \, m_j^1 \, \delta_{sj} = i \frac{M_N^4}{2 v_\chi} \, \delta_{sp} \, , \\
\sum_{j} \tilde{\Gamma}_{spj}^{1,0,0} &= i \frac{M_N}{2 v_\chi} \sum_{k,r=4}^6  \left( \bar{M}^\dagger \right)_{sr} \delta_{kj} \,\delta_{pj} = i \frac{M_N^2}{2 v_\chi} \, \delta_{sp} \, , \\
\sum_{j} \tilde{\Gamma}_{spj}^{1,1,0} &= i \frac{M_N}{2 v_\chi} \sum_{k,r=4}^6  \left( \bar{M}^\dagger \bar{M} \bar{M}^\dagger \right)_{sr} \delta_{kj} \, \delta_{pj} = i \frac{M_N^4}{2 v_\chi} \, \delta_{sp} \, .
\end{align}
With these results one can directly obtain from \eqref{eq:LRR} and \eqref{eq:LRRtilde}
\begin{align}
\left( L_\eta^{RR} \right)_{sp}^{(0)} = - \frac{i}{2 v_\chi} \left( f_5 \, M_N^2 + f_6 \, M_N^4 \right) \delta_{sp} \, , \quad \left( \tilde{L}_\eta^{RR} \right)_{sp}^{(0)} = - \frac{i}{2 v_\chi} \left( f_9 \, M_N^2 + f_{10} \, M_N^4 \right) \delta_{sp} \, .
\end{align}
Using the $f_{i}$ and $F_{i,j,...,n}$ loop functions
defined in Appendix~\ref{sec:app2}, we find
\begin{align}
\left(L_\eta\right)^{\beta \alpha} = i \, M_\ell \, y \, \frac{ M_N^2}{2 v_\chi \left( M_N^2 - m_\eta^2 \right)^2}  \left( M_N^2- m_\eta^2 + m_\eta^2 \log \frac{m_\eta^2}{M_N^2} \right) y^\dagger
\end{align}
and we recover the known result
of~\cite{Babu:2007sm,Escribano:2021ymx}~\footnote{The definition of
the $y$ Yukawa matrix in \cite{Escribano:2021ymx} differs from ours by
a complex conjugation.}
\begin{equation}
  - \mathcal{L}_{J\ell\ell} =\frac{iJ}{16 \pi^2 v_\chi} \bar{\ell}\left( M_\ell \, y \,  \Theta \,  y^\dagger  \, P_L -   y  \, \Theta \,  y^\dagger \, M_\ell \, P_R \right) \ell \, ,
\end{equation}
with
\begin{equation}
  \Theta_{sp} \equiv \frac{M_N^2}{\left( M_N^2 - m_\eta^2 \right)^2}  \left( M_N^2- m_\eta^2 + m_\eta^2 \log \frac{m_\eta^2}{M_N^2} \right) \delta_{sp} \, .
\end{equation}
It is important to emphasize here that calculating this coupling in
this general form in this model is an overkill. In this model, as
explained above, the majoron only interacts with the fermion singlets
and the coupling can be taken to be diagonal without loss of
generality. This fact implies that the resulting loop integral is
straightforward, since the light neutrinos do not enter the
loop. Therefore, if we were to perform the explicit calculation, we
would obtain a simple integral that would directly yield the
reproduced result. In other words, we would not need to carry out any
expansion. Hence, the fact that our calculation, after considering the
most general possible model and carrying out all expansions while
accounting for all possibilities, returns the correct result is a very
strong validation of our outcome.

Finally, one can estimate the size of the majoron couplings to charged
leptons in the Scotogenic model. In this model, the coupling $S$ can
be estimated as
\begin{equation} \label{eq:Sscoto}
  S_{\rm Scot} \sim \frac{M_\ell \, y^2}{16 \pi^2 \, v_\chi} \sim \frac{2 \, M_\ell \, m_\nu}{\lambda_5 \, v^2} \sim \frac{32 \pi^2}{\lambda_5} \, S_{\rm Type-I} \, ,
\end{equation}
where we have used that $m_\nu \sim \lambda_5 \, v^2 y^2 / (32 \pi^2
M_N)$, with $\lambda_5$ the coupling of the quartic scalar term
$\left( H^\dagger \eta \right)^2$, and $M_N \sim v_\chi$ in this
model. We have also assumed that all loop functions are of
$\mathcal{O}(1)$ and have compared to the result for the type-I seesaw
in Eq.~\eqref{eq:StypeI}. We find a large enhancement compared to the
tiny value obtained in the type-I seesaw. In fact, a relatively small
$\lambda_5$ coupling would naturally lead to the violation of the
bounds in Sec.~\ref{sec:coup}. For instance, the current limit on
BR$(\mu \to e \, J) \sim 10^{-17}/\lambda_5^2$ would be violated for
$\lambda_5 \lesssim 10^{-6}$. This result confirms previous findings
in the literature, which have already shown a very rich majoron
phenomenology in the Scotogenic model. We refer to
\cite{Escribano:2021ymx} for a detailed exploration of the
phenomenology of the $\ell_\alpha \to \ell_\beta \, J$ decays in this
model.

\subsection{Type-I seesaw with two Higgs doublets}
\label{subsec:typeI2HDM}

{
\renewcommand{\arraystretch}{1.4}
\begin{table}[tb]
\centering
{\setlength{\tabcolsep}{0.5em}
\begin{tabular}{c c c c c c c}
\toprule  
field & spin & generations & $\mathrm{SU(3)}_c$ & $\mathrm{SU(2)}_L$ & $\mathrm{U(1)}_Y$ & \U1L \\
\midrule   
$N$ & $\frac{1}{2}$ & 3 & \one & \one & 0 & 1 \\
\midrule
$\chi$ & 0 & 1 & \one & \one & 0 & -2 \\
\midrule
$H_2$ & 0 & 1 & \one & \two & $\frac{1}{2}$ & 0 \\
\bottomrule
\end{tabular}
}
\caption{New particles in the type-I seesaw with two Higgs doublets and spontaneous lepton number violation.}
\label{tab:typeI2HDM}
\end{table}
}

\begin{figure}
\centering
  \includegraphics[width=0.6\linewidth]{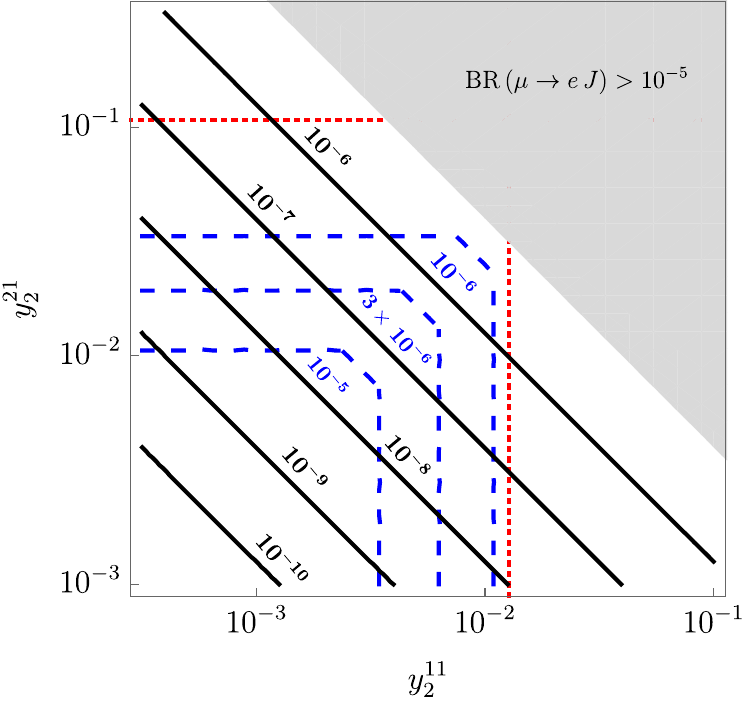}
\caption{Contours of BR($\mu \to e \, J$) (black) and $\lambda_5^{\rm
    max}$ (blue, dashed) in the $y_2^{11}$-$y_2^{21}$ plane. The rest
  of $y_2$ Yukawa couplings are set to zero for simplicity, while
  $\lambda = 1$ and $v_\chi = 1$ TeV. Here $\lambda_5^{\rm max}$ is
  the maximal value of $\lambda_5$ compatible with all the entries in
  $m_\nu^{\rm 1-loop}$ being smaller than those in $m_\nu^{\rm
    tree}$. The shaded region is excluded by $\mu \to e \, J$ searches
  while the vertical and horizontal red dotted lines determine the
  upper limits on $y_2^{11}$ and $y_2^{21}$, respectively, derived
  from Eqs.~\eqref{eq:See} and \eqref{eq:Smumu}.
    \label{fig:2HDM}
    }
\end{figure}

In order to better illustrate our results, we now consider a
non-minimal model: the type-I seesaw with two Higgs doublets and
spontaneous lepton number violation. This model extends the Two Higgs
Doublet Model (2HDM) particle content with 3 singlet fermions $N$ and
a scalar singlet $\chi$, all charged under the global \U1L as shown in
Tab.~\ref{tab:typeI2HDM}, just as the type-I seesaw discussed in
Sec.~\ref{subsec:typeI}. In addition, this model adds a second Higgs
doublet, $H_2$. In the following we assume that we are working in the
Higgs basis, in which $H_2$ does not acquire a VEV. The Yukawa terms
relevant for our discussion are
\begin{equation}
    -\mathcal{L} =Y \, \bar{L} H e_R + Y_2 \, \bar{L} H_2 e_R + y \, \bar{L} \tilde{H} N + y_2 \, \bar{L} \tilde{H_2} N + \frac{1}{2} \, \lambda \, \chi \, \bar{N}^c N + \hc \, . 
\end{equation}
Since we work in the Higgs basis, $M_D$ and $M_R$ are given by the
same expressions as in Eq.~\eqref{eq:MDMR}. Similarly, the charged
leptons mass matrix is
\begin{equation}
  M_\ell = Y \, \frac{v}{\sqrt{2}} \, .
\end{equation}
Furthermore, $Y_2$ and $y_2$ are two general $3 \times 3$
matrices. Neutrino mass generation takes place in exactly the same way
as in the minimal type-I seesaw and the same holds true for the
majoron. Consequently, the gauge diagrams remain identical to those we
computed previously in Sec.~\ref{subsec:typeI}. However, as in any
2HDM, the presence of a second Higgs doublet implies the existence of
physical charged and CP-odd scalars. Following the same notation as in
Sec.~\ref{sec:coup}, they will be denoted as $\eta$ and
$\sigma$. These states induce the Feynman diagrams with amplitudes
$\M_\eta$ and $\M_{\sigma n}$ shown in Figs.~\ref{fig:EtaRhodiag} and
\ref{fig:Sigmadiag}. In order to compute them, we first identify the
new couplings, not present in Sec.~\ref{subsec:typeI}, defined in
Eqs.~\eqref{eq:rotA}-\eqref{eq:rotF}. These are
\begin{align}
  \bar{B}_{kr} &= \left\{ \begin{array}{cl}
    \displaystyle i \left(y_2\right)_{kr} \, , & \text{if} \,\, k = 1,2,3 \, \, \text{and} \, \, r = 4,5,6 \\
     & \\
    \displaystyle 0 \, , & \text{otherwise} \end{array} \right. \\
  C^{\alpha \beta} &= - \left( Y_2 \right)_{\beta \alpha}^* \, , \\
  \bar{D}_L^{\alpha k} &= \left\{ \begin{array}{cl}
    \displaystyle \left(Y_2\right)_{\alpha k} \, , & \text{if} \,\, k  = 1,2,3 \\
     & \\
    \displaystyle 0 \, , & \text{otherwise} \end{array} \right. \\
  \bar{D}_R^{\alpha k} &= \left\{ \begin{array}{cl}
    \displaystyle \left(y_2\right)_{\alpha k} \, , & \text{if} \,\, k  = 4,5,6 \\
     & \\
    \displaystyle 0 \, , & \text{otherwise} \end{array} \right.
\end{align}
In our computation we will assume that $m_\eta,m_\sigma \ll M_H$. Let
us start with the computation of the charged scalar contribution. We
have to obtain expressions for $L_\eta^{RR}$, $\tilde{L}_\eta^{RR}$,
$L_\eta^{LL}$, $\tilde{L}_\eta^{LL}$ and $L_\eta^{RL}$. We will start
by computing the first order of $L_\eta^{RR}$ and
$\tilde{L}_\eta^{RR}$. One can easily find
\begin{align}
\sum_{j} \Delta_{spj}^{0,1,1} = \sum_{j} \tilde{\Gamma}_{spj}^{1,1,0} =  \frac{i}{2 v_\chi} M_R^4 \, \delta_{sp} + \O(M_R^{2})  \, , \\ 
\sum_{j \sim h} \Delta_{spj}^{1,1,-1} = \sum_{j } \tilde{\Gamma}_{spj}^{1,0,0} =  \frac{i}{2 v_\chi} M_R^2 \, \delta_{sp} + \O(M_R^{0})  \, ,
\end{align}
while other terms are of lower order in $M_R$ and hence
subdominant. With these results one can directly obtain
\begin{align}
\left( L_\eta^{RR} \right)_{sp}^{(0)} = - \frac{i}{2 v_\chi} \left( f_5 \, M_R^2 + f_6 \, M_R^4 \right) \, \delta_{sp}  \, , \quad \left( \tilde{L}_\eta^{RR} \right)_{sp}^{(0)} = - \frac{i}{2 v_\chi} \left( f_9 \, M_R^2 + f_{10} \, M_R^4 \right) \, \delta_{sp} \, .
\end{align}
It is now useful to notice that $f_i \sim M_H^{-2}$ for $i$ odd and $f_i
\sim M_H^{-4}$ for $i$ even. Therefore, the two terms in the previous
expressions are of the same order in $M_R$. In fact, making use of the
expressions for the $f_i$ and $F_{i,j,...,n}$ loop functions in
Appendix~\ref{sec:app2}, and using $m_\eta \ll M_H$, we find
\begin{align}
\left( L_\eta^{RR} \right)_{sp}^{(0)} = \frac{-3 i}{8 v_\chi} \, \delta_{sp}  \, , \quad  \left( \tilde{L}_\eta^{RR} \right)_{sp}^{(0)} = \frac{-i}{8 v_\chi} \, \delta_{sp} \, ,
\end{align}
where we have used that $\frac{\left(M_R\right)_{sp}}{M_H} =
\delta_{sp}$ at first order. Our result is easy to understand: since
both the majoron and the charged scalar couple directly to the
singlets $N$, which are essentially the heavy neutrinos, there is no
mixing in the loop (no mass suppression) at leading order. For this
reason, one expects $L_\eta^{LL}$ and $\tilde{L}_\eta^{LL}$ to be
subdominant, as two mixings are required in the loop. This fact is
reflected in our results and an explicit calculation confirms this
reasoning. Moreover, one could be tempted to make a similar argument
for $L_\eta^{RL}$, which requires one mixing. However, we must be
careful with this term because it is not proportional to the charged
lepton masses in Eq.~\eqref{eq:Leta}. In fact, the introduction of
$M_D$ is equivalent to the charged lepton mass, since $M_D \sim
M_\ell$. Let us see it explicitly. One can easily find
\begin{align}
- \sum_{j} \left( \tilde{\Gamma}_{psj}^{0,2,0} \right)^* = \sum_{j} \tilde{\Delta}_{spj}^{1,1,1}=\sum_{j} \Delta_{spj}^{1,1,1} = \sum_{j\sim l} \left( \Gamma_{psj}^{0,2,0} \right)^* = \frac{i}{2 v_\chi} \left( M_R^4 M_D^\dagger \right)_{s-3, \, p} + \O(M_R^{2})  \, , \\
\sum_{j \sim h} \Delta_{spj}^{0,1,-1} =\sum_{j } \tilde{\Delta}_{spj}^{1,0,1} = - \sum_{j } \left( \tilde{\Gamma}_{psj}^{0,1,0} \right)^* =  \sum_{j \sim l} \left( \Gamma_{psj}^{0,1,0} \right)^* =\frac{i}{2 v_\chi} \left( M_R^2 M_D^\dagger \right)_{s-3, \, p} + \O(M_R^{0})   \, ,
\end{align}
while other terms are of lower order in $M_R$ and can be
neglected. With these results,
\begin{align}
\left( L_\eta^{RL} \right)_{sp}^{(0)} = -2 \frac{i}{2 v_\chi} \left[ \left( f_3 \, M_R^2 + f_4 \, M_R^4 \right)M_D^\dagger \right]_{s-3, \, p} \, ,
\end{align}
and, using the expressions for $f_i$ and $F_{i,j,...,n}$, we obtain
\begin{align}
\left( L_\eta^{RL} \right)_{sp}^{(0)} = - \frac{i}{2 v_\chi} \left( M_D^\dagger \right)_{s-3, \, p} \, .
\end{align}
And, with both results, finally,
\begin{align}
\left( L_\eta \right)^{\beta \alpha} = \frac{i}{2 v_\chi} \left( M_\ell \, y_2 \,  y_2^\dagger + Y_2 \,  M_D \, y_2^\dagger \right)_{\beta \alpha} \, .
\end{align}\\
Let us continue with the CP-odd scalar contribution, $\M_{\sigma
  n}$. First, we note that the divergent piece vanishes, as
expected. This is because only the real part contributes, see
Eq.~\eqref{eq:Lsigman1}, and $\left( L_{\sigma n}
\right)^\textup{(div)}$ turns out to be purely imaginary. To show it
we compute $\sum_j \Delta_{psj}^{1,0,1}$, $\sum_j
\Gamma_{spj}^{0,1,0}$ and $ \sum_j \tilde{\Gamma}_{spj}^{0,1,0}$,
\begin{align}
\sum_j \Delta_{psj}^{1,0,1} &= \frac{i}{2 v_\chi} \left( M_R^2 M_D^\dagger \right)_{p-3, \, s} \, , \\
\sum_j \tilde{\Gamma}_{spj}^{0,1,0} &=  \frac{i}{2 v_\chi}  \left( M_D M_R^2 \right)_{p-3, \, s} \, , \\
\sum_j \Gamma_{spj}^{0,1,0} &= 0 \, .
\end{align}
Therefore, 
\begin{align}
\left( L_{\sigma n} \right)^\textup{(div)} &= \text{Re} \, \Bigg\{ -\frac{1}{2 v_\chi} \left[ \textup{Tr} \left( y_2 M_R^2 M_D^\dagger \right) - \textup{Tr} \left( y_2^* M_R^2 M_D^T \right) \right] \left( 3 + 2\, \frac{1}{\epsilon} + 2\,  \log \frac{\mu^2}{M_H^2}\right) \Bigg\} = 0 \, .
\end{align}
We then consider the finite piece. We have already computed several
terms required for the evaluation of $L_\eta^{RL}$. With these results
(and a few more), one can show that $L_{\sigma n}^{(0)}$ vanishes for
the same reason as for the divergent piece. This means that the
leading order contribution from the $\sigma$ scalar is, at most, of
order one. This includes the explicit order one contribution given in
Appendix~\ref{sec:app1} as well as the \textit{hidden} order one
contribution in $L_{\sigma n}^{(0)}$. However, one can show that this
order also cancels out, so $L_{\sigma n} = \mathcal{O}(M_R^{-2})$, and
in this way, we can safely neglect this contribution. In summary, the
majoron coupling to a pair charged leptons in the type-I seesaw with
two Higgs doublets is given by
\begin{align}
\mathcal{L}_{\ell\ell J} =\frac{iJ}{16\pi^2 v_\chi} \bar{\ell}\Bigg\{&M_\ell ~ \textup{Tr} \left( \frac{M_D M_D^\dagger }{v^2} \right) \gamma_5 +2 \left(M_\ell \frac{M_D M_D^\dagger}{v^2} -M_\ell \, y_2 \,  y_2^\dagger - Y_2 \,  M_D \, y_2^\dagger \right) P_L \nonumber \\
&-2  \left( \frac{M_D M_D^\dagger}{v^2} M_\ell -   y_2 \,  y_2^\dagger \, M_\ell - y_2 \,  M_D^\dagger \, Y_2^\dagger \right) P_R \Bigg\} \ell \, .
\end{align}
Let us now study the size of the coupling $S$ in this model. It can be
split as
\begin{equation}
  S_{\rm Type-I+2HDM} = S_{\rm Type-I} + S_{\rm 2HDM} \, ,
\end{equation}
where $S_{\rm Type-I}$ is given in Eq.~\eqref{eq:StypeI} and $S_{\rm
  2HDM}$ contains the new terms associated to our 2HDM version of the
model, not present in the type-I seesaw (with one Higgs doublet). The
latter can be estimated as
\begin{equation} \label{eq:Type-I2HDM}
  S_{\rm 2HDM} \sim \frac{1}{16 \pi^2 \, v_\chi} \left( M_\ell \, y_2^2 + y_2 \, Y_2 \, M_D \right) \sim \frac{1}{16 \pi^2 \, v^2 \, v_\chi} \left( M_\ell \, M_D^{' 2} + M_\ell' \, M_D \, M_D' \right) \, ,
\end{equation}
where we have defined $M_\ell'$ and $M_D'$ analogously to $M_\ell$ and
$M_D$, simply replacing $Y$ and $y$ by $Y_2$ and $y_2$. This allows us
to obtain the ratio
\begin{equation}
  \frac{S_{\rm 2HDM}}{S_{\rm Type-I}} = \left( \frac{M_D'}{M_D} \right)^2 + \frac{M_\ell' \, M_D'}{M_\ell \, M_D} \, .
\end{equation}
This result hints at an enhacement of the majoron coupling to charged
leptons compared to the type-I seesaw if $M_D' \gg M_D$ and/or
$M_\ell' \gg M_\ell$. Let us suppose $M_\ell' = 0$ (or, equivalently,
$Y_2 = 0$) and consider $M_D' \gg M_D$ (or, equivalently, $y_2 \gg
y$). One should note that $y_2$ does not participate in the type-I
seesaw contribution to neutrino masses, but it does induce a
Scotogenic-like 1-loop contribution, $m_\nu^{\rm 1-loop} \sim
\lambda_5 \, v^2 y_2^2 / (32 \pi^2 M_R)$.~\footnote{In the absence of
mixing between the CP-even scalars, $\lambda_5$ is the coupling of the
quartic scalar term $\left( H^\dagger H_2 \right)^2$. Otherwise, if
the CP-even components of $H^0$ and $H_2^0$ mix, $\lambda_5$ is a
combination of scalar potential parameters associated to this mixing.}
Imposing $m_\nu^{\rm 1-loop} \lesssim m_\nu \equiv m_\nu^{\rm tree}$
then leads to
\begin{equation} \label{eq:2HDMhier}
  \frac{\lambda_5}{32 \pi^2} \left( \frac{y_2}{y} \right)^2 \lesssim 1 \, .
\end{equation}
Assuming now $y \sim 10^{-6}$, the typical size of the Yukawa
couplings for a type-I seesaw at the TeV scale,
Eq.~\eqref{eq:2HDMhier} implies $\lambda_5 \, y_2^2 \lesssim
10^{-10}$. This is fulfilled for $y_2 \sim 1$ if $\lambda_5 \lesssim
10^{-10}$. In this case, $M_D'/M_D \sim 10^6$ and one expects an
enhancement of the majoron couplings of about $12$ orders of magnitude
with respect to the type-I seesaw. Such a huge increase would indeed
lead to conflict with the bounds in Eqs.~\eqref{eq:See},
\eqref{eq:Smumu}, \eqref{eq:mulim} and \eqref{eq:taulim}. However,
less pronounced hierarchies lead to observable effects, for instance
in $\ell_\alpha \to \ell_\beta \, J$ decays, in agreement with the
current constraints. This is shown in Fig.~\ref{fig:2HDM}, which
displays contours of BR($\mu \to e \, J$) and $\lambda_5^{\rm max}$ in
the $y_2^{11}$-$y_2^{21}$ plane. Here $\lambda_5^{\rm max}$ is the
maximal value of $\lambda_5$ compatible with all the entries in
$m_\nu^{\rm 1-loop}$ being smaller than those in $m_\nu^{\rm
  tree}$. The rest of $y_2$ Yukawa couplings are set to zero for
simplicity, while $\lambda = 1$ and $v_\chi = 5$ TeV. This plot
clearly shows that the model can achieve large $\mu \to e \, J$
braching ratios, almost as large as the current bound, in the region
of parameter space that has been considered. In conclusion, the
extension of the type-I seesaw with a second Higgs doublet induces
novel contributions to the majoron couplings to charged leptons that
may increase them notably and lead to observable consequences.

\section{Final discussion}
\label{sec:sum}

The majoron is present in any model that breaks a global lepton number
symmetry spontaneously. In the absence of explicit sources of \U1L
breaking, its mass is exactly zero. This has a strong impact on the
phenomenology of the model, which changes dramatically with respect to
the variant with explicit breaking. We have derived general analytical
expressions for the 1-loop coupling of the majoron to charged
leptons. These couplings are known to be crucial for the phenomenology
of the model and, in fact, they often lead to the most stringent
constraints. Our analytical results can be used in virtually all
models featuring a majoron. We have provided several example models
that illustrate how to use them. Due to our lack of imagination, we
could not find examples for all possible contributions, but they can
be present in some (less popular) models. In any case, we found
perfect agreement with the results obtained in all models for which
the majoron coupling to charged leptons was known. This is a clearly
non-trivial cross-check of our analytical expressions.

The main limitation of our approach is the assumption of a hierarchy
of mass scales, which is nevertheless common to most Majorana neutrino
mass models, such as those based on the seesaw mechanism. We have also
focused on scenarios that actually \textit{require} the computation of
the 1-loop majoron couplings to charged leptons. For instance, since
we did not consider additional fermions in the particle spectrum, one
may wonder about the applicability of our results in models with heavy
leptons. While our analytical expressions cannot be directly applied
in this case, we note that (i) the analogous expressions with heavy
neutral leptons can be easily adapted, and, more importantly, (ii)
models with heavy charged leptons induce majoron couplings with the
light charged leptons already at tree-level, thus making our 1-loop
results mere corrections. In conclusion, our analytical expressions
are applicable to any model in which the 1-loop majoron couplings to
charged leptons \textit{must} be computed.

Our results go beyond the majoron. One can use them to obtain the
1-loop couplings of any massless (or very light) pseudoscalar to a
pair of charged leptons. In fact, the majoron can be regarded as a
particular and theoretically well-motivated type of axion-like
particle (ALP). Whenever the assumptions of our approach are
fulfilled, our results can also be applied to compute the 1-loop
couplings of an ALP to a pair of charged leptons. Therefore, they may
be of interest in scenarios not necessarily related to neutrino mass
generation and lepton number violation.

The study of majorons and lepton number violation represents a
fascinating frontier in particle physics, pushing the boundaries of
our understanding of the fundamental building blocks of the
Universe. Our contribution constitutes just another step in the
ongoing quest to detect and characterize majorons. As we delve deeper
into the mechanisms behind neutrino masses and lepton number
violation, we not only expand our knowledge of the SM but also pave
the way for potential breakthroughs that could revolutionize our
understanding of particle interactions and the origins of mass.

\section*{Acknowledgements}

The authors are grateful to Salvador Centelles Chuli\'a for
discussions. Work supported by the Spanish grants PID2020-113775GB-I00
(AEI/10.13039/501100011033) and CIPROM/2021/054 (Generalitat
Valenciana). The work of AHB is supported by the grant No. CIACIF/2021/100 (also funded by Generalitat Valenciana).  AV acknowledges financial support from MINECO through the Ramón y Cajal contract
RYC2018-025795-I.

\appendix

\section{Higher order terms}
\label{sec:app1}

We present here our results for the contributions of higher order in
$m_{\rm light}$ not included in Sec.~\ref{sec:coup}, using exactly the
same notation and conventions:
\begin{align}
  \left( \Gamma_Z \right)^{(1)} &= - \sum_{s=1}^3  \sum_{i,j\sim l} \bar{A}_{kr} U_{ki} U_{rj}\log \left(\frac{m_j^2}{M_H^2}\right) \left(U_{si}m_i U_{sj}^*+U_{sj}m_j U_{si}^*\right) \, ,
\end{align}
\begin{align}
  \left( L_W^{\beta \alpha} \right)^{(1)} &=  \sum_{i,j \sim l} \bar{A}_{kr}^* U_{\beta i}^* m_i U_{ki}^* U_{\alpha j} U_{rj}^* \left[ -\frac{5}{6} +3\log \left(\frac{M_H}{M_W} \right)+2\log \left(\frac{m_i}{M_H} \right)\right] \nonumber \\
  & - \sum_{i,j \sim l} \bar{A}_{kr} U_{\alpha i} m_i U_{ki} U_{\beta j}^* U_{rj} \left[ \frac{7}{6} +2\log \left(\frac{m_i}{M_H} \right)\right]
\, ,
\end{align}
\begin{align}
  \left(L_\eta^{LL}\right)^{(1)} =& \left(L_\eta^{RR}\right)^{(1)} \left( m_i \leftrightarrow m_j, k \leftrightarrow r, s \leftrightarrow p \right) \, , \\
  \left(L_\eta^{RR}\right)^{(1)}_{sp} =&  \sum_{j\sim l} \bar{A}_{kr} U_{rj}m_j U_{sj} \Bigg[ \left(F_{5,7,-1} -f_{13} \right)\delta_{kp}-\left(F_{8,2}+ f_{14} \right) \left( \bar{M}^\dagger \bar{M} \right)_{kp} \nonumber \\
    +& \sum_{i\sim l}U_{ki}U_{pi}^*\frac{\log \frac{m_j}{m_\eta}}{m_\eta^2} \Bigg] \, , \\
  \left( L_\eta^{RL} \right)^{(1)}_{sp} =& \sum_{j\sim l}\left[  F_{3,11,-1} \left(\tilde{\Delta}_{s p}^{1,0,1}+\Delta_{s p}^{1,0,1}\right)+F_{4,12,-2} \, \tilde{\Delta}_{s p}^{1,1,1} + F_{12,-2} \, \Delta_{s p}^{1,1,1}  \right]  \, ,
\end{align}
\begin{align}
  \left( L_{\sigma n} \right)^{(1)} &= \text{Re} \, \Bigg\{ \sum_{j\sim l} \left[ \bar{B}_{sp}\left(- \frac{2}{M_H^2} \Delta_{ps}^{1,1,1}+\Delta ^{1,0,1}_{ps}+ \tilde{\Delta}_{ps}^{1,0,1} \right) \right] \Bigg\} \, , \\
  \left( L_{\sigma n} \right)^{(2)} &= \text{Re} \, \Bigg\{ \sum_{i,j \sim l} \bar{A}_{kr}U_{rj}U_{ki}\left[\bar{B}_{sp}U_{pj}U_{si} m_j m_i+\bar{B}_{sp}^* U_{sj}^* U_{pi}^*\left(m_i^2 + m_j^2 \right)\right]\log \frac{M_H^2}{m_j^2} \Bigg\} \, .
\end{align}

\section{Loop functions}
\label{sec:app2}

We define the following loop functions, which turn out to be relevant
for the charged $\eta$ scalar contributions:
\begin{align}
  f_1 &= \frac{-M_H^2 + m_{\eta}^2 + (2 M_H^2 - m_{\eta}^2) \log\left(\frac{M_H^2}{m_{\eta}^2}\right)}{2(M_H^2 - m_{\eta}^2)^2} \, ,
\end{align}
\begin{align}
f_2 &= -\frac{-M_H^2 + m_{\eta}^2 + M_H^2 \log\left(\frac{M_H^2}{m_{\eta}^2}\right)}{2 M_H^2(M_H^2 -  m_{\eta}^2)^2} \,,  \\
f_3 &= \frac{2 M_H^2 (M_H^2 - m_{\eta}^2) + (-3 M_H^2 m_{\eta}^2 + m_{\eta}^4) \log\left(\frac{M_H^2}{m_{\eta}^2}\right)}{2 (M_H^2 - m_{\eta}^2)^3} \, , \\
f_4 &= \frac{-M_H^4 + m_{\eta}^4 + 2 M_H^2 m_{\eta}^2 \log\left(\frac{M_H^2}{m_{\eta}^2}\right)}{2 M_H^2 (M_H^2 - m_{\eta}^2)^3} \, , \\
f_5 &= \frac{6 M_H^6 - 5 M_H^4 m_{\eta}^2 - 2 M_H^2 m_{\eta}^4 + m_{\eta}^6 - 2 (6 M_H^4 m_{\eta}^2 - 4 M_H^2 m_{\eta}^4 + m_{\eta}^6) \log\left(\frac{M_H^2}{m_{\eta}^2}\right)}{4 (M_H^2 - m_{\eta}^2)^4} \, , \\
f_6 &= \frac{-3 M_H^6 - 2 M_H^4 m_{\eta}^2 + 7 M_H^2 m_{\eta}^4 - 2 m_{\eta}^6 + (8 M_H^4 m_{\eta}^2 - 2 M_H^2 m_{\eta}^4) \log\left(\frac{M_H^2}{m_{\eta}^2}\right)}{4 M_H^2 (M_H^2 - m_{\eta}^2)^4} \, , \\
f_7 &= -\frac{-4 M_H^8 - 19 M_H^6 m_{\eta}^2 + 27 M_H^4 m_{\eta}^4 - 5 M_H^2 m_{\eta}^6 + m_{\eta}^8 + 6 M_H^4 m_{\eta}^2 (3 M_H^2 + m_{\eta}^2) \log\left(\frac{M_H^2}{m_{\eta}^2}\right)}{6 (M_H^2 - m_{\eta}^2)^5} \, , \\
f_8 &= \frac{-M_H^6 - 9 M_H^4 m_{\eta}^2 + 9 M_H^2 m_{\eta}^4 + m_{\eta}^6 + 6 M_H^2 m_{\eta}^2 (M_H^2 + m_{\eta}^2) \log\left(\frac{M_H^2}{m_{\eta}^2}\right)}{3 (M_H^2 - m_{\eta}^2)^5} \, , \\
f_9 &= \frac{2 M_H^6 - 11 M_H^4 m_{\eta}^2 + 10 M_H^2 m_{\eta}^4 - m_{\eta}^6 + (8 M_H^2 m_{\eta}^4 - 2 m_{\eta}^6) \log\left(\frac{M_H^2}{m_{\eta}^2}\right)}{4 (M_H^2 - m_{\eta}^2)^4} \, , \\
f_{10} &= -\frac{M_H^6 - 6 M_H^4 m_{\eta}^2 + 3 M_H^2 m_{\eta}^4 + 2 m_{\eta}^6 + 6 M_H^2 m_{\eta}^4 \log\left(\frac{M_H^2}{m_{\eta}^2}\right)}{4 M_H^2 (M_H^2 - m_{\eta}^2)^4} \, , \\
f_{11} &=\frac{2 M_H^6 + 3 M_H^4 m_\eta^2 - 6 M_H^2 m_\eta^4 + m_\eta^6 - 
 6 M_H^4 m_\eta^2 \log \frac{ M_H^2}{ m_\eta^2}}{4\left(M_H^2 - m_\eta^2 \right)^4} \, ,\\
 f_{12} &=-\frac{ M_H^4 + 4 M_H^2 m_\eta^2 - 5 m_\eta^4 -4M_H^2 m_\eta^2 \log \frac{ M_H^2}{ m_\eta^2} - 
 2  m_\eta^4 \log \frac{ M_H^2}{ m_\eta^2}}{4\left(M_H^2 - m_\eta^2 \right)^4} \, \\
f_{13} &= -\frac{3 M_H^4 - 4 M_H^2 m_{\eta}^2 + m_{\eta}^4 + 2 M_H^4 \log\left(\frac{m_{\eta}^2}{M_H^2}\right)}{2 (M_H^2 - m_{\eta}^2)^3} \, , \\
f_{14} &= \frac{2 (M_H^2 - m_{\eta}^2) + (M_H^2 + m_{\eta}^2) \log\left(\frac{m_{\eta}^2}{M_H^2}\right)}{2 (M_H^2 - m_{\eta}^2)^3} \, ,
\end{align}
\begin{align}
f_{15} &= \frac{M_H^8 - 8 M_H^6 m_{\eta}^2 + 8 M_H^2 m_{\eta}^6 - m_{\eta}^8 - 12 M_H^4 m_{\eta}^4 \log\left(\frac{m_{\eta}^2}{M_H^2} \right)}{6 \left (M_H^2-m_\eta^2\right )^5 } \, , \\
f_{16} &=\frac{-M_H^6 + 9 M_H^4 m_\eta^2 + 9 M_H^2 m_\eta^4 - 17 m_\eta^6 + 6 (3 M_H^2 m_\eta^4 + m_\eta^6) \log\left(\frac{m_\eta^2}{M_H^2}\right)}{12 (M_H^2 - m_\eta^2)^5} \, .
\end{align}
In addition, we define
\begin{align}
F_{i,j,k} \equiv \textup{sign}(i) f_i + \textup{sign}(j) f_j + \textup{sign}(k) f_k \, .
\end{align}
The functions in this Appendix take simplified expressions if $m_\eta
\ll M_H$ and $m_\eta \gg M_H$. In the light $\eta$ limit ($m_\eta \ll
M_H$) they take the approximate expressions
\begin{align}
f_1 &= \frac{1}{2M_H^2} \left[-1 + 2\log \left(\frac{M_H^2}{m_\eta^2}\right) \right] \, , \\
f_2 &= -\frac{1}{2M_H^4} \left[-1 +\log \left(\frac{M_H^2}{m_\eta^2}\right) \right] \, , \\
f_{13} &= -\frac{1}{2M_H^2} \left[3 -2 \log \left(\frac{M_H^2}{m_\eta^2}\right) \right] \, , \\
f_{14} &= \frac{1}{2M_H^4} \left[2 - \log \left(\frac{M_H^2}{m_\eta^2}\right) \right] \, , \\
f_3 &= \frac{2}{3} f_5 = \frac{3}{2} f_7 = 2 f_9 = 2 f_{11} = 6 f_{15} = \frac{1}{M_H^2} \, , \\
f_4 &= \frac{2}{3} f_6 = \frac{3}{2} f_8 = 2 f_{10} = 2 f_{12} = 6 f_{16} = -\frac{1}{2 M_H^4} \, , 
\end{align}
whereas in the heavy $\eta$ limit ($m_\eta \gg M_H$) they can be
approximately written as
\begin{align}
f_1 &= -\frac{1}{2 m_\eta^2} \left[ -1 + \log \left(\frac{M_H^2}{m_\eta^2}\right) \right] \, , \\
f_3 &= -\frac{1}{2m_\eta^2} \log \left(\frac{M_H^2}{m_\eta^2}\right) \, , \\
f_5 &= \frac{1}{4 m_\eta^2} \left[ 1 - 2\log \left(\frac{M_H^2}{m_\eta^2}\right) \right] \, , \\
f_9 &= -\frac{1}{4 m_\eta^2} \left[1 + 2\log \left(\frac{M_H^2}{m_\eta^2}\right) \right] \, , \\
f_{12} &= \frac{1}{4 m_\eta^2 M_H^2} \left[5+2 \log \left(\frac{M_H^2}{m_\eta^2}\right)\right] \, ,
\end{align}
\begin{align}
f_{14} &= -\frac{1}{2 m_\eta^4} \left[-2 + \log \left(\frac{m_\eta^2}{M_H^2}\right) \right] \, , \\
f_{16} &= \frac{1}{12 m_\eta^4} \left[17 - 6\log \left(\frac{m_\eta^2}{M_H^2}\right) \right] \, , \\
f_2 &= f_4 = f_6 = f_{10} = -\frac{1}{2 m_\eta^2 M_H^2} \, , \\
3 f_7 &= 2 f_{11} = f_{13} = 3 f_{15} = \frac{1}{2 m_\eta^2} \, , \\
f_8 &= -\frac{1}{3 m_\eta^4} \, .
\end{align}

\bibliographystyle{utphys}
\bibliography{refs}

\end{document}